\documentclass[12pt]{article}
\usepackage{a4,amssymb,verbatim,amsmath}

\usepackage[cp866]{inputenc}

\newtheorem{theorem}{Theorem}[section]
\newtheorem{lemma}[theorem]{Lemma}
\newtheorem{remark}[theorem]{Remark}
\newtheorem{corollary}[theorem]{Corollary}
\newtheorem{proposition}[theorem]{Proposition}

\newtheorem{ex}{Example}[section]

\newtheorem{ass}{Assumption}[section]

\numberwithin{equation}{section}

\DeclareMathOperator{\sgn}{sgn}

\begin{document}

\newcommand{\ddt}{\partial \over \partial t}
\newcommand{\ddx}{\partial \over \partial x}
\newcommand{\ddy}{\partial \over \partial y}

\def\dhp#1{ \mathop{#1}\limits_{h}}

\newcommand{\ufrac}[8]{\frac{(u_{#1} - u_{#2})(u_{#3} - u_{#4})}{(u_{#5} - u_{#6})(u_{#7} - u_{#8})} }
\newcommand{\ufracc}[8]{\frac{(u_{#1} - u_{#2})(u_{#3} - u_{#4})}{(u_{#5} - u_{#6})(u_{#7} - u_{#8})^2} }

\begin{center}
{ \Large {\bf First integrals of ordinary difference  equations  \\
beyond Lagrangian methods.}  \\
 $ $ \\
A pedagogical treatment.}
\end{center}

\bigskip

\begin{center}
{\bf V. Dorodnitsyn}$^{{*}}$,
{\bf  E. Kaptsov}$^{**}$,
{\bf R. Kozlov}$^{\dag}$,
{\bf  P. Winternitz}${{\ddag}}$
\end{center}

$^{{*}}$
Keldysh Institute of Applied Mathematics of Russian Academy of Science,
Miusskaya Pl. 4, Moscow, 125047, Russia;
{e-mail: dorod@spp.Keldysh.ru}

$^{**}$
OAO Magnit, Solnechnaya st. 15/5, Krasnodar, 350072, Russia;
{e-mail: evgkaptsov@gmail.com}

$^{\dag}$
Department of Business and Management Science,
Norwegian School of Economics,
Helleveien 30, 5045, Bergen, Norway;
{e-mail: Roman.Kozlov@nhh.no}

${{\ddag}}$
Centre de Recherches Math\'ematiques et D\'epartement de math\'ematiques et de statistique,
Universit\'e de Montr\'eal,
Montr\'eal, QC, H3C 3J7, Canada;
{e-mail: wintern@crm.umontreal.ca}

\bigskip

\begin{center}
{\bf 07.11.2013}
\end{center}

\bigskip

\begin{center}
{\bf \Large Abstract }
\end{center}

A new method for finding first integrals of discrete equations is
presented. It can be used for discrete equations which do not
possess a variational (Lagrangian or Hamiltonian) formulation. The
method is based on a newly established identity which links
symmetries of the underlying discrete equations, solutions of the
discrete adjoint equations and first integrals. The method is
applied to invariant mappings and discretizations of a second order and a third
order ODEs. In examples the set of independent first integrals makes it possible to find the
general solution of the discrete equations. The method is compared to a
direct method of constructing first integrals.

\eject

\tableofcontents

\eject

\section{Introduction}

\label{intr}

Considerable progress has been made over the last 25 years in the
applications of Lie group theory to difference equations (for
reviews see~\cite{[D-book], [LW-2], [LW-3]}
and for original papers~\cite{[Pavel], [Pavel2], [Dor_3], [Dor_4], [Dor_5], [Dheat], [D_K_W_1],
[Hydon-Mansfield], [LW-4], [LW-5], [LW-6], [LW-11], [LW-12], [Olver-1], [Quispel-1]}).
The overall aim of the program is to turn Lie group theory into a tool 
for solving discrete equations that is as efficient as it is for differential ones.


For ordinary differential equations (ODEs) one of the important
applications of Lie group theory is to reduce the order of the
equation and ideally to solve it analytically and explicitly.
Essentially there are two ways of doing this, once a nontrivial Lie
point symmetry group of the equation is found. One is to perform a
transformation of the independent and dependent variables that takes
the Lie algebra into a convenient form. This also transforms the
equation to a form in which the reduction of the order becomes
obvious.

An alternative method is to use the Lie point symmetry group to
construct first integrals of the equation that are of lower order
than the equation (or system of equations) itself. This can be done
if a Lagrangian exists and the symmetries are variational ones. If a
sufficient number of first integrals can be obtained using the
Noether theorem, then the derivatives can be eliminated from the
set of first integrals. This provides a solution of the original equation by purely
algebraic operations, without any changes of variables or any
integration.

If no invariant Lagrangian exists alternative methods of
constructing lower order first integrals have been proposed
in~\cite{[Bluman1], [Bluman2]}, and in~\cite{[Ibr2], [ibr11a], [Ibr]}. They make
use of the so called adjoint equation solutions of which one uses
to construct the required first integrals. We shall call this the
"adjoint equation method".

The integration methods based on transformations of coordinates have
not been adapted to difference equations. The algebraic methods
based on invariant Lagrangians and Hamiltonians have been adapted
and successfully applied to solve three point difference schemes
in~\cite{[66], [D-book], [DKap],  [D_K_W_1],  [45]} and~\cite{[DK-1], [DK-2], [DK-3]}, respectively.
A research note on adapting the "adjoint
equation method" to difference equations has been published
in~\cite{[DAN]}.

The purpose of this article is to present and justify the adjoint
equation method for difference systems with an arbitrary number of
variables and also to document its usefulness on examples. The paper
is organized as follows.
In Section~\ref{adjointsection} we present  a brief summary of
the adjoint equation method for an arbitrary system of partial or
ordinary differential equations (PDEs or ODEs).
Section~\ref{sectionODE} specializes  the theory sketched
in Section~\ref{adjointsection} to the case of one scalar ODE.
Section~\ref{integrating} describes integrating factors for scalar ODEs.
The adjoint equation method for discrete systems is presented in Section~\ref{discretetheory}.
This theory is specialized to the case of scalar discrete  equations (mappings)
and discretizations of scalar ODEs in Sections~\ref{discrete}
and~\ref{discretization}, respectively. Section~\ref{integrating2}
presents integrating factors for scalar discrete equations. Finally,
Section~\ref{conclusion} provides concluding remarks.

\section{Adjoint equation method for constructing conservation laws for differential equations}

\label{adjointsection}

Let us consider  a system of $n$-th order PDEs
\begin{equation}  \label{equat}
F _ {\beta}   ( {\bf x}   ,  {\bf u} ,
{ \mathop{\bf u}\limits_{1}}  ,  { \mathop{\bf u}\limits_{2}} ,..., { \mathop{\bf u}\limits_{n}} ) = 0  ,
\qquad  {\beta} = 1, ..., r ,
\end{equation}
where $ {\bf x}  = ( x^1 , ... , x^p ) $, $ {\bf u}  = ( u^1 , ... , u^q ) $,
$$
{ \mathop{\bf u}\limits_{1}} := \{ u_i ^k  \}
=  \left\{   {\partial u ^k  \over \partial {x^i}  } \right\} ,
\quad
... ,
\quad
{ \mathop{\bf u}\limits_{s}} := \{ u_{i_1 ... i_s} \}
=  \left\{   {\partial ^s u ^k  \over \partial x^ {i_1}   ... \partial x^ {i_s} } \right\} ,
\quad
... ,
$$
$ i = 1, ..., p $, $k = 1, ..., q$.

Let $L_{ \alpha \beta  }$ be a linear operator
\begin{equation}  \label{sym}
L _{ \alpha \beta  }= \sum _{k = 0 } ^{\infty}
F _{ \beta ,  u ^{\alpha}  _{i_1...i_k} }   D_{i_1} \ldots  D_{i_k}  ,
\qquad
F _{ \beta ,  u ^{\alpha}  _{i_1...i_k} }
= { \partial  F _{ \beta } , \over \partial u ^{\alpha}  _{i_1...i_k} }  ,
\end{equation}
where
$$
D_i= {\partial {}\over \partial {x^i}}
+ u ^k _{i}{\partial {}\over \partial {u ^k }}
+ v ^k _{i}{\partial {}\over \partial {v ^k }}
+ u ^k _{ij}{\partial {}\over \partial {u ^k _j}}
+ v ^k _{ij}{\partial {}\over \partial {v ^k _j}}
+ u ^k _{ijl}{\partial {}\over \partial {u ^k _{jl}}}
+ v ^k _{ijl}{\partial {}\over \partial {v ^k _{jl}}}
+ ...,
$$
then the {\it adjoint equations} are given by the variational derivatives
(or Euler-Lagrange operators):
\begin{equation}  \label{adjointi}
F^*  _{\alpha} =  L ^*  _{ \alpha \beta }   v ^{ \beta }
=  {\delta {} \over \delta  u ^{ \alpha }  } ( v ^{ \beta }   F _{ \beta }  )
= \sum _{k = 0 } ^{\infty}   (-1)^k D_{i_1} \ldots  D_{i_k}
(  v ^{ \beta }  F_{ \beta , u ^{\alpha }  _{ i_1...i_k} } )  = 0 ,
\quad
\alpha  = 1, ..., q .
\end{equation}
We assume summation over repeated indices.
Notice that the adjoint equations are always linear equations for
${\bf v} = ( v^1, ..., v^r ) $
with coefficients that in general depend upon {\bf u} (solution of~(\ref{equat})).

The {\bf basic operator identity}  is the following
\begin{equation}  \label{ident0}
v ^{\beta}   L  _{\alpha \beta }  w ^{\alpha}   - w ^{\alpha}   L^* _{\alpha \beta }  v ^{\beta}    = D_i C^i,
\end{equation}
where $v ^{\beta}  $  and $w ^{\alpha}  $ are some functions of ${\bf x}$, ${\bf u}$ and derivatives of ${\bf u}$.
Here
\begin{equation}  \label{conslaw}
C^i
= \sum _{k = 0 } ^{\infty}   D_{i_1} \ldots  D_{i_k} ( w ^{\alpha} )
{\delta {} \over \delta  u ^{\alpha}  _{i i_1...i_k} }  ( v ^{\beta}  F _{\beta}  )  ,
\end{equation}
where
$$
{\delta {} \over \delta  u ^ {\alpha}   _{i i_1...i_k}  }
=   \sum _{s = 0 } ^{\infty}   (-1)^s   D_{i_1} \ldots  D_{i_s}
{ \partial  \over \partial u ^{\alpha}  _{ i i_1...i_k     i_1...i_s} }
$$
are higher order variational operators (or higher order
Euler-Lagrange operators). Since the scalar ($ q = r = 1$)
relation is probably due to Lagrange  (see for example~\cite{[Dem]},
Eq.~(2.75) on p.~80), we refer to~(\ref{ident0}) as the { \it
Lagrange identity}.


We will be interested in Lie symmetries ~\cite{[Ovs1], [Ibr1], [Olver1]}
\begin{equation}  \label{symmetr}
X =  \xi  ^i  {\partial \over \partial  x ^i  }
+   \eta ^{\alpha}  {\partial \over \partial  u ^{\alpha}  }
+  \sum _{s = 1} ^{\infty}  \zeta ^{\alpha} _{i_1 ... i_s }  {\partial \over \partial  u ^{\alpha}  _{i_1 ... i_s }   }  ,
\end{equation}
where $\xi  ^i $ and $   \eta ^{\alpha}  $
are some functions of ${\bf x} $, ${\bf u}$
and a finite number of derivatives of ${\bf u}$ and
$$
\zeta ^{\alpha} _{i_1 ... i_s } = D _{i_1}  ... D_{ i_s } (  \eta ^{\alpha} - \xi ^i u _i ^{\alpha}  ) + \xi ^i u  _{i   {i_1 ... i_s } }  ^{\alpha} .
$$
Note that for point symmetries $\xi  ^i $ and $   \eta ^{\alpha}  $
depend only on  ${\bf x} $ and ${\bf u}$.
To each symmetry~(\ref{symmetr})
there corresponds the canonical or evolutionary symmetry
\begin{equation}    \label{symcanon}
\bar{X}
=   \bar{\eta} ^{\alpha}   {\partial \over \partial  u  ^{\alpha} }
+  \sum  _{s = 1} ^{\infty}
\bar{\zeta} ^{\alpha} _{i_1 ... i_s }  {\partial \over \partial  u  ^{\alpha} _{i_1 ... i_s }   }  ,
\end{equation}
 where
$$
\bar{\eta} ^{\alpha}  = \eta ^{\alpha}  - \xi ^i u _i ^{\alpha}   ,
\qquad
\bar{\zeta} _{i_1 ... i_s }  ^{\alpha}  =  D _{i_1}  ... D_{ i_s }   (    \bar{\eta} ^{\alpha}  )  .
$$

The identity~(\ref{ident0}) can be used to link
symmetries of the differential equations~(\ref{equat}),
solutions of the corresponding adjoint equations~(\ref{adjointi})
and conservation laws.

Choosing  $w ^{\alpha} = \bar{\eta} ^{\alpha}  =  \eta ^{\alpha}   - \xi ^i  u_i ^{\alpha}    $,
we obtain the identity
\begin{equation}  \label{idet55}
v ^{\beta}  \bar{X} F _ {\beta}  =  \bar{\eta} ^{\alpha}   F^* _ {\alpha}   +  D_i C^i
\end{equation}
for evolutionary operators~(\ref{symcanon}). For the Lie symmetry operator ~(\ref{symmetr}) we obtain
\begin{equation}   \label{idet56}
v  ^{\beta}   X  F _ {\beta}
= v ^{\beta}  \xi  ^i  D_i  F _{\beta}
+   ( \eta ^{\alpha}  - \xi ^i u _i ^{\alpha} )  F^*  _{\alpha}
+ D_i C^i.
\end{equation}
Here the quantities $C^i$ are
\begin{equation}  \label{conslaw2}
C^i
= \sum _{k = 0 } ^{\infty}  D_{i_1} \ldots  D_{i_k} ( \eta ^{\alpha}  - \xi ^i u _i ^{\alpha}  )
{\delta {} \over \delta  u ^{\alpha}  _{i i_1...i_k} }  ( v ^{\beta}  F _{\beta}  )  .
\end{equation}

The following theorem is based on the Lagrange
identity:

\begin{theorem}
The system of equations~(\ref{equat}) and its adjoint system~(\ref{adjointi}) possess the
following conservation law
\begin{equation}  \label{cons}
\left.
{ D_i C^i   }
\right|_{(\ref{equat}),(\ref{adjointi})}  = 0
\end{equation}
for each Lie symmetry~(\ref{symmetr}) of the differential equation~(\ref{equat})
and for each solution of the adjoint equation~(\ref{adjointi}).
\end{theorem}

\noindent {\it Proof.}
The result follows directly from Eq.~(\ref{idet56}). Indeed, $X F _ {\beta}   = 0$
because it is a symmetry criterion for Eq.~(\ref{equat}),
$D_i ( F _{\beta}  ) =0$ since it is a differential consequence of Eq.~(\ref{equat}),
and $F^*  _{\alpha} =0$ on a solution of adjoint equations~(\ref{adjointi}).
\hfill $\Box$
\medskip

Since we are interested in Eqs.~(\ref{equat}) we need
conservation laws for these equations alone, i.e., without using
solutions of the adjoint equations~(\ref{adjointi}).

One can  get rid of the adjoint variables ${\bf v}  $
figuring in the conservation law~(\ref{cons}) and subsequent formulas.
The identity~(\ref{idet55})
and the idea of solving the adjoint equations in terms of solutions of the original equations
were explicitly presented in  \cite{[Bluman1]} (see also \cite{[Bluman2]}).
These ideas were   also suggested  and further developed
in \cite{[Ibr2], [ibr11a], [Ibr]},
where  numerous examples for ODEs and PDEs were worked out explicitly.
The introduction of adjoint variables, of linear equations
adjoint to nonlinear ones and the extension variational principles for
equations without classical Lagrangians were also considered
in~\cite{[AthertonHomsy], [KaraMahomed], [Rosenhaus1]} and others.

\begin{theorem}    \label{main0}
Let the adjoint equations~(\ref{adjointi}) be satisfied
for all solutions of the differential equations~(\ref{equat})
upon a substitution
\begin{equation}   \label{subst0}
{\bf v} = {\bf \varphi } ( {\bf x}   ,  {\bf u },
{ \mathop{\bf u}\limits_{1}} , { \mathop{\bf u}\limits_{2}} , ...  ) ,
\qquad
{\bf \varphi }    {\not\equiv}   { \bf  0 }.
\end{equation}
Then, any Lie symmetry~(\ref{symmetr}) of the equations~(\ref{equat})
leads to the conservation law
\begin{equation}  \label{cons00}
\left.
{ D_i C^i   }
\right|_{(\ref{equat})}  = 0 ,
\end{equation}
where ${\bf v}$ and its derivatives should be eliminated via equation~(\ref{subst0})
and its differential consequences.
\end{theorem}

\begin{remark}
Equations~(\ref{equat}) and~(\ref{adjointi}) can be considered as variational equations
for the formal Lagrangian~\cite{[Ibr]}
\begin{equation}   \label{formality}
{\cal L} = v ^{\beta}  F _ {\beta}  ,
\end{equation}
which provides the original and adjoint equations
\begin{equation}
F _  {\beta}  = {  \delta {\cal L}  \over \delta   v  ^{\beta} }  = 0 ,
\quad
 {\beta} = 1, ..., r,
\qquad
F^* _{\alpha} = {  \delta {\cal L}  \over \delta   u ^{\alpha} }  = 0 ,
\quad
\alpha = 1, ..., q .
\end{equation}
\end{remark}

\begin{remark}
The same operator identities~(\ref{idet55})  and~(\ref{idet56})  form the basis of
the Noether theorem~\cite{[Noe1]} for the Lagrangian systems  (see~\cite{[Ibr1]} for details).
Indeed, consider the case $r=1$,  put
 $v = 1$ and apply it to a Lagrangian
$F = {\cal L}  ( {\bf x}  ,  {\bf  u } ,  { \mathop{\bf u}\limits_{1}}  ,  { \mathop{\bf u}\limits_{2}} ,... )$.
Then we get the following identities
$$
\bar{X}  {\cal L}
= \bar{\eta}^{\alpha}   {\delta   {\cal L} \over \delta u^{\alpha}  }
+ D_i ( \bar{N}^i   {\cal L} ),
$$
$$
 \bar{N}^i = \sum _{k = 0 } ^{\infty}  D_{i_1} \ldots  D_{i_k} (  \bar{\eta}  ^{\alpha} )
{\delta {} \over \delta  u ^{\alpha}  _{i i_1...i_k} }
$$
and
$$
X  {\cal L}  +  {\cal L}   D_i  \xi ^i
 = ( \eta ^{\alpha}     - \xi ^i  u_i ^{\alpha}    )       {\delta {\cal L} \over \delta u^{\alpha}  }
+ D_i   (  N ^i  {\cal L} ),
$$
$$
 N^i =  \xi ^i
+ \sum _{k = 0 } ^{\infty}  D_{i_1} \ldots  D_{i_k} (       \eta ^{\alpha}   - \xi ^i  u_i ^{\alpha}  )
{\delta {} \over \delta  u ^{\alpha}  _{i i_1...i_k} }   ,
$$
which yield a conservation law
$$
D_i \bar{C} ^i  = 0 , \qquad  \bar{C} ^i = \bar{N}^i  {\cal L}
$$
and
$$
D_i  C ^i  = 0 , \qquad  C ^i =  N ^i   {\cal L} ,
$$
correspondingly,
for the  Euler-Lagrange equations
$$
{  \delta {\cal L}  \over \delta   u ^{\alpha} }  = 0 ,
\qquad
\alpha = 1, ..., q .
$$
Operators $N^i$ and $\bar{N}^i$ are called Noether operators.
\end{remark}

\section{The case of one ordinary differential equation }

\label{sectionODE}

In this section we restrict ourselves to scalar ordinary differential equations.
It is a particular case of the general theory sketched in the previous section.
We restrict ourselves to Lie point symmetries
because later we will consider the discrete case, to which we wish to adapt the Lie point symmetry approach.

\subsection{The case of an n-th order ODE}

\label{ODEtheory}

Let us consider a scalar ODE of order n
\begin{equation}  \label{ode}
F  ( x,  u  ,  \dot{u}   , \ddot{u}   , ... , u^{(n)}   ) = 0  .
\end{equation}
We will be interested in Lie point symmetries
\begin{equation}  \label{symmetry}
X =  \xi(x,u)  {\partial \over \partial  x  }
+   \eta(x,u)  {\partial \over \partial  u  }
+  \zeta _1   {\partial \over \partial  \dot{u}   }
+  \zeta _2  {\partial \over \partial  \ddot{u}   }
+ ...
+  \zeta _k   {\partial \over \partial  u^{(k)}  }
+ ... ,
\end{equation}
where
$$
\zeta _k = D ^k (  \eta - \xi \dot {u} ) + \xi u^{(k+1)}
$$
and
\begin{equation*}
D =  {\partial \over \partial  x  }
+  \dot{u}   {\partial \over \partial  u  }
+  \dot{v}   {\partial \over \partial  v  }
+  \ddot{u}   {\partial \over \partial  \dot{u}  }
+  \ddot{v}   {\partial \over \partial  \dot{v}  }
+ ...
+  u^{(k+1)}   {\partial \over \partial  u^{(k)}    }
+  v^{(k+1)}   {\partial \over \partial  v^{(k)}    }
+ ...
\end{equation*}
is the total differentiation.

To each Lie point symmetry~(\ref{symmetry})
 there corresponds the  symmetry in evolutionary form
\begin{equation}
 \bar{X}
 =   \bar{\eta}   {\partial \over \partial  u  }
 +  \bar{\zeta} _1   {\partial \over \partial  \dot{u}   }
 +  \bar{\zeta} _2  {\partial \over \partial  \ddot{u}   }
  + ...
 +  \bar{\zeta} _k   {\partial \over \partial  u^{(k)}  }
 + ... ,
\end{equation}
 where
\begin{equation*}
 \bar{\eta}  = \eta (x,u) -  \xi(x,u) \dot{u}   ,
\end{equation*}
\begin{equation*}
\bar{\zeta} _1 = D   (    \bar{\eta} )  ,
\qquad
... ,
\qquad
\bar{\zeta} _k = D ^k  (    \bar{\eta} )  .
\end{equation*}

By means of the variational operator
\begin{equation}   \label{euler}
{  \delta  \over \delta   u  }
=     {\partial  \over \partial  u  }
- D    {\partial   \over \partial  \dot{u}  }
+ D ^2     {\partial   \over \partial  \ddot{u}  }
+ ...
+  (-1) ^{k}  D^k    {\partial   \over \partial  u^{(k)}    }
+ ...
\end{equation}
we introduce the adjoint equation
\begin{equation} \label{adjoint}
F^* = {  \delta  \over \delta   u  } ( v F ) = 0.
\end{equation}
Thus~(\ref{adjointi}) simplifies  to
\begin{equation*}
F^* =  v     {\partial F \over \partial  u  }
- D  \left( v   {\partial  F \over \partial  \dot{u}  } \right)
+ D ^2  \left( v   {\partial  F \over \partial  \ddot{u}  } \right)
+ ...
+  (-1) ^{n}  D^n   \left( v  {\partial  F \over \partial  u^{(n)}    } \right)
= 0 .
\end{equation*}

\begin{remark}
Equations~(\ref{ode}) and~(\ref{adjoint}) can be considered as variational equations
for the formal Lagrangian~\cite{[Ibr]}
\begin{equation}   \label{formal}
{\cal L } = v F ,
\end{equation}
which provides the original and adjoint equations
\begin{equation}
F = {  \delta L  \over \delta   v  }  = 0 ,
\qquad
F^* = {  \delta L  \over \delta   u  }  = 0 .
\end{equation}
\end{remark}

Let us define higher order  variational (or Euler--Lagrange) operators
\begin{equation} \label{higher}
{  \delta  \over \delta   u ^{(i)}  }
=     {\partial  \over \partial  u^{(i)}  }
- D    {\partial   \over \partial   u^{(i+1)} }
+ D ^2     {\partial   \over \partial  u^{(i+2)}  }
+ ...
+  (-1) ^{k}  D^k    {\partial   \over \partial  u^{(i+k)}    }
+ ...
\end{equation}
Note that Euler--Lagrange operator~(\ref{euler}) belongs to this set as
$$
{  \delta  \over \delta   u  }
= {  \delta  \over \delta   u ^{(0)}  }  .
$$

\begin{lemma}  \label{mainindentity}  {\bf (Main identity for scalar ODEs)}
The following identity holds
\begin{equation}     \label{identity2}
v    X   F   =   v \xi D F  +  ( \eta  -  \xi \dot{u} )   F^*   + D  I    ,
\end{equation}
where
\begin{equation}
I
= \sum_{i = 0} ^{n-1} D^i ( \eta  -  \xi \dot{u} ) {\delta \over \delta u ^{(i+1)} } ( v F ) .
\end{equation}
\end{lemma}

\noindent This is a special case of~(\ref{idet56}), (\ref{conslaw2}).

\medskip

We prefer identity~(\ref{identity2}) instead of the corresponding identity
for the canonical operator
\begin{equation}   \label{canon}
v    \bar{X}  F   =  \bar{\eta}  F^*   + D  I    .
\end{equation}
In the discrete case the framework of Lie point symmetries is much better developed
in terms of standard vector fields~(\ref{symmetry}) than evolutionary ones.
The goal of this paper is to develop a discrete analog of the identity~(\ref{identity2}).





Now we examine the identity~(\ref{identity2})  on the solutions of the ODE~(\ref{ode}).
The left hand side turns out to be zero if operator $X$ is a symmetry of the ODE.
The first term on the right hand side contains $D F $
and  drops out as a differential consequence of the ODE.
We are left with
\begin{equation}
 \left.  ( \eta  -  \xi \dot{u} )   F^*  \right| _{F = 0 } +   \left.   D I   \right| _{F = 0 }  = 0 .
\end{equation}
If we can find a substitution for   function $v$ providing $  F^*  = 0 $,
then we can get rid off the adjoint equation.
Thus, we obtain a first integral of the ODE.
Let us formulate this as the following theorem.

\begin{theorem}   \label{main} {\bf (Main theorem for scalar ODEs)}
Let the adjoint equation~(\ref{adjoint}) be satisfied
for all solutions of the original ODE~(\ref{ode})
upon a substitution
\begin{equation}  \label{subst}
v = \varphi ( x,u ) ,
\qquad
\varphi    {\not\equiv} 0 .
\end{equation}
Then, any Lie point symmetry~(\ref{symmetry}) of the equation~(\ref{ode})
leads to first integral
\begin{equation}  \label{firstint1}
I
= \sum_{i = 0} ^{n-1} D^i  ( \eta  -  \xi \dot{u} )  {\delta \over \delta u ^{(i+1)} } ( v F )  ,
\end{equation}
where $v$ and its derivatives should be eliminated via equation~(\ref{subst})
and its differential consequences.
\end{theorem}

\begin{remark}
First integrals $I$, given by~(\ref{firstint1}), can depend on $u^{(n)}$ as well as higher derivatives.
We will call such expression {\it higher} first integrals.
It is reasonable to use the ODE~(\ref{ode}) and its differential consequences
to express these first integrals as functions of the minimal set of variables,
i.e.,  in the form $ \tilde{I}  (x, u, \dot{u} , ..., u^{(n-1)} )$.
In the examples of this section we will bypass the higher first integrals  $I$
and provide only the final results $ \tilde{I}$.
\end{remark}

\begin{remark} Theorem~\ref{main} can be extended
from point-wise substitutions~(\ref{subst})
to differential substitutions
\begin{equation}   \label{substit}
v = \varphi ( x,u , \dot{u} ),
\qquad
... ,
\qquad
v = \varphi ( x,u , \dot{u}, \ddot{u} , ... , u ^{(n-1)} ),
\qquad
\varphi   {\not\equiv} 0  .
\end{equation}
\end{remark}


\subsection{Second order ODEs}

\label{Secondorder}

The method of the adjoint equation also works for differential equations
which possess a Lagrangian. We demonstrate its application
 for the case of a second order ODE
$$
F  ( x,  u  ,  \dot{u}   , \ddot{u}     ) = 0 .
$$
Its adjoint equation~(\ref{adjoint}) is
$$
F^* =  v     {\partial F \over \partial  u  }
- D  \left( v   {\partial  F \over \partial  \dot{u}  } \right)
+ D ^2  \left( v   {\partial  F \over \partial  \ddot{u}  } \right)
= 0  .
$$
First integrals~(\ref{firstint1}) are given by the formula
$$
I = \left[
( \eta - \dot{u} \xi )
\left(  {\partial   \over \partial  \dot{u}  }
- D {\partial   \over \partial  \ddot{u}  } \right)
+ D ( \eta - \dot{u} \xi )     {\partial   \over \partial  \ddot{u}  }
\right] ( v F )
$$

\bigskip

\noindent {\bf Example: Harmonic oscillator}

\medskip

Let us consider the one-dimensional harmonic oscillator
\begin{equation}   \label{ode2}
 F = \ddot{u}  + u = 0 .
\end{equation}
It admits the symmetries~\cite{[Lut]}
\begin{equation*}
X_1 = {\frac{ \partial }{\partial x}} ,
\qquad
X_2 = \sin x  {\frac{ \partial }{\partial u}} ,
\qquad
X_3 = \cos x {\frac{ \partial }{\partial u}} ,
\qquad
X_4 = u {\frac{ \partial }{\partial u}} ,
\end{equation*}
\begin{equation}  \label{harnomicsym}
X_5 = \sin 2x  {\frac{ \partial }{\partial x}}  +  u  \cos 2x  {\frac{ \partial }{\partial u}}   ,
\qquad
X_6 = \cos 2x  {\frac{ \partial }{\partial x}}  -  u  \sin 2x  {\frac{ \partial }{\partial u}}   ,
\end{equation}
\begin{equation*}
X_7 = u \sin x  {\frac{ \partial }{\partial x}}  +  u^2   \cos x  {\frac{ \partial }{\partial u}}  ,
\qquad
X_8 = u \cos x  {\frac{ \partial }{\partial x}}  -  u^2   \sin x  {\frac{ \partial }{\partial u}} .
\end{equation*}

The adjoint equation~(\ref{adjoint}) takes the form
\begin{equation}
F^* = \ddot{v} + v = 0 .
\end{equation}
Note that the equation~(\ref{ode2}) is {\it self-adjoint}:
$$
\left. F^* \right| _{ v =  u}  = F .
$$
Let us choose the solution $v(x,u) = u $,
then for symmetries~(\ref{harnomicsym}) we obtain the first integrals
$$
\tilde{I}_1 =  \dot{u}^2  + u^2 ,
\qquad
\tilde{I}_2 =   - \dot{u}   \sin  x +  u \cos  x   ,
\qquad
\tilde{I}_3 =  - \dot{u}  \cos x  - u \sin x    ,
$$
$$
\tilde{I}_4 \equiv 0  ,
\qquad
\tilde{I}_5 = ( \dot{u}^2  - u^2 )  \sin 2x  -  2 u \dot{u}   \cos 2x  ,
$$
$$
\tilde{I}_6 = ( \dot{u}^2  - u^2 ) \cos 2x   +  2 u \dot{u}   \sin  2x  ,
\qquad
\tilde{I}_7  \equiv 0   ,
\qquad
\tilde{I}_8  \equiv 0  ,
$$
respectively.
Choosing values of two independent first integrals $\tilde{I}_2 = A $ and $\tilde{I}_3 = B $,
we obtain the well-known general solution as
$$
u(x) = A  \cos x - B  \sin x  .
$$

\subsection{Third order ODEs}

\label{mainexample}

To the third order ODE
$$
F  ( x,  u  ,  \dot{u}   , \ddot{u} , \dddot{u}    ) = 0
$$
there corresponds the adjoint equation
$$
F^* =  v     {\partial F \over \partial  u  }
- D  \left( v   {\partial  F \over \partial  \dot{u}  } \right)
+ D ^2  \left( v   {\partial  F \over \partial  \ddot{u}  } \right)
- D ^3  \left( v   {\partial  F \over \partial  \dddot{u}  } \right)
= 0  .
$$
First integrals are given as
\begin{multline*}
I = \left[
( \eta - \dot{u} \xi )
\left(  {\partial   \over \partial  \dot{u}  }
- D {\partial   \over \partial  \ddot{u}  }
+ D^2  {\partial   \over \partial  \dddot{u}  }
\right)
\right.
\\
\left.
+ D ( \eta - \dot{u} \xi )
\left(   {\partial   \over \partial  \ddot{u}  }
- D   {\partial   \over \partial  \dddot{u}  }
\right)
+ D ^2 ( \eta - \dot{u} \xi )  {\partial   \over \partial  \dddot{u}  }
\right] ( v F )  .
\end{multline*}

\bigskip

\noindent {\bf Example}

\medskip

Let us investigate the ODE
\begin{equation}  \label{third00}
 F =  { 1 \over \dot{u} ^2 } \left( \dot{u}  \dddot{u} - { 3 \over 2 }  \ddot{u} ^2  \right) - f(x)  = 0 .
\end{equation}
Its numerical solutions were considered in~\cite{[Pavel], [Pavel2]}
using a symmetry-preserving discretization.

The first term is the well-known Schwarzian derivative that has many
interesting and important applications in mathematics, physics and
even (originally) in cartography
(for an interesting review see~\cite{[Ovsienko]}).

In the general case this ODE admits the symmetry group $SL(2, \mathbb{R} )$.
Its Lie algebra is realized as
\begin{equation}   \label{part00}
X_1 = {\frac{ \partial }{\partial u}} ,
\qquad
X_2 = u {\frac{ \partial }{\partial u}} ,
\qquad
X_3 = u^2 {\frac{ \partial }{\partial u}} ,
\qquad
\end{equation}
for $f=M= \mbox{const}$ we get an additional symmetry
\begin{equation}   \label{part01}
X_4 = {\frac{ \partial }{\partial x}},
\end{equation}
and for $f=M = 0 $ there are two further symmetries
\begin{equation}  \label{part02}
X_5 = x {\frac{ \partial }{\partial x}} ,
\qquad
X_6 = x^2 {\frac{ \partial }{\partial x}} .
\end{equation}

The ODE that we shall consider is
\begin{equation}  \label{third}
 F =  { 1 \over \dot{u} ^2 } \left( \dot{u}  \dddot{u} - { 3 \over 2 }  \ddot{u} ^2  \right) - M  = 0 ,
 \qquad M = \mbox{const} .
\end{equation}
The equation~(\ref{third}) can be integrated by standard integration techniques.
Indeed, it can be rewritten as
$$
\left( { 1 \over \sqrt{|\dot{u} |} } \right) '' + { M \over 2} {
1 \over \sqrt{|\dot{u} |} } = 0
$$
and solved for $  { 1 \over \sqrt{|\dot{u} |} }  $ as a linear
(Schr\"{o}dinger) equation.
We find  the solution $u(x)$ for
different cases of the parameter $M$ as follows
\begin{equation}    \label{sol1}
M = 0 : \qquad
u (x) = { 1 \over C_1 x + C_2 } +  C_3
\quad
\mbox{or}
\quad
u (x) =  C_1 x + C_2 ;
\end{equation}
\begin{multline}    \label{sol2}
M < 0 :
\qquad
u (x) =  C_1 \tanh ( \omega x + C_2 ) + C_3 \\
\mbox{or}
\quad
u (x) =  C_1 \coth ( \omega x + C_2 ) + C_3 \\
\mbox{or}
\quad
u (x) = C_1 e^{ \pm 2 \omega  x }  + C_2 ,
\qquad
\omega = \sqrt{ - { M / 2}} ;
\end{multline}
\begin{equation}    \label{sol3}
M > 0 :
\qquad
u (x) = C_1 \tan ( \omega x + C_2 ) + C_3 ,
\qquad
\omega = \sqrt{  { M / 2}} ;
\end{equation}
where $C_1 \neq 0 $, $C_2$ and $C_3$  are integration constants.

\begin{remark}
Let us mention that the restriction $C_1 \neq 0 $ can be removed
if instead of the ODE~(\ref{third}) we consider the equation
$$
 \dot{u} ^2 F
 =    \dot{u}  \dddot{u} - { 3 \over 2 }  \ddot{u} ^2  - M \dot{u} ^2  = 0  .
$$
\end{remark}

Unfortunately, this method will not work in the discrete case, so a
different approach is needed and will be based on a discrete version
of Theorem~\ref{main}.
Let us first  solve ODE~(\ref{third}) using Theorem~\ref{main}.
The idea is to find three independent first integrals of~(\ref{third})
and then to eliminate the derivatives $\dot{u}$ and $\ddot{u}$ from them
(third order ODEs can have at most three first integrals).

The adjoint equation~(\ref{adjoint}) takes the form
\begin{equation}  \label{adjoint0}
F^* = - { 1 \over \dot{u} }  (  \dddot{v}  + 2 M    \dot{v} )   =   0 .
\end{equation}

We can consider different cases for the substitutions~(\ref{subst})
and~(\ref{substit}).

\bigskip

\noindent { \bf Ansatz~1.}
Let us look for  solutions of the form $v =  v (x) $,
which is the simplest Ansatz.
It also directly provides the general solution of the adjoint equation.
We obtain three independent solutions of the adjoint equation~(\ref{adjoint0})
\begin{equation}   \label{indep1}
\begin{array}{lllll}
M = 0 :
 & v_a = 1  ,  & v_b = x, & v_c  = x^2 ;  &  \\
M < 0 : & v_a = 1  ,
 & v_b = \cosh ( 2\omega  x )  , & v_c  = \sinh ( 2 \omega x ) ,
 &  \omega = \sqrt{ - { M / 2}}   ; \\
M  > 0 :
 & v_a = 1  ,  & v_b = \cos ( 2 \omega  x )  , & v_c  = \sin (  2 \omega x ) ,
 & \omega = \sqrt{ M / 2}  .  \\
\end{array}
\end{equation}
We will use these solutions of the adjoint equation to find first
integrals of the ODE~(\ref{third}).

\bigskip

Let us use symmetries~(\ref{part00}),(\ref{part01}),(\ref{part02})
and solutions of the adjoint equation~(\ref{adjoint0}) to construct first integrals.
The notation  $\tilde{I}_{j \alpha }$ means that this integral
corresponds to symmetry $X_j$ and solution $v_{\alpha}$ of the adjoint equation.

For all values of the parameter $M$
there is only one common solution of the adjoint equation, namely
\begin{equation}    \label{328a}
v_a (x) = 1 .
\end{equation}
It provides us with the first integrals
\begin{multline}  \label{common1}
\tilde{I}_{1a}
=   { 1\over 2}  { \ddot{u}^2 \over \dot{u}^3 }
+ {M \over \dot{u} }  ,
\qquad
\tilde{I}_{2a}
= u \left(
{ 1\over 2} { \ddot{u}^2 \over \dot{u}^3 }
+  { M  \over \dot{u} }
\right)
- { \ddot{u} \over \dot{u} }  ,
\\
\tilde{I}_{3a}
= u^2 \left(
{ 1\over 2}  { \ddot{u}^2 \over  \dot{u}^3 }
+ { M   \over \dot{u} }
\right)
- 2 u  {   \ddot{u} \over \dot{u} }
+ 2 \dot{u}  ,
\qquad
\tilde{I}_{4a}
\equiv  - 2M .
\end{multline}
Two additional first integrals for $M = 0$ are trivial:
\begin{equation}
\tilde{I}_{5a} \equiv  0 ,
\qquad
\tilde{I}_{6a} \equiv  -2 .
\end{equation}

The non-trivial first integrals obey the relation
$$
\tilde{I}_{1a} \tilde{I}_{3a} - \tilde{I}_{2a}  ^2 = 2 M  .
$$
Thus we have only two independent first integrals
and it is not sufficient  for the integration of the third order ODE.
To find a sufficient number of first integrals
we will consider solutions of the adjoint equation
which are specific for particular cases of the parameter $M$.
Let us go through different cases of the parameter.

\bigskip

\noindent {\bf Case:  $M = 0$}

\medskip

The additional solutions of  the adjoint equation are
\begin{equation*}  \label{special1}
v _b (x)  = x
\qquad
\mbox{and}
\qquad
v _c (x) = x^2  .
\end{equation*}
For $ v _b (x)  = x$ we obtain first integrals
\begin{multline}    \label{firstint1b}
\tilde{I} _{1b}
= { x \over 2} { \ddot{u}^2 \over  \dot{u}^3 }
+ { \ddot{u} \over  \dot{u}^2 } ,
\qquad
\tilde{I}_{2b}
= u \left(
{ x \over 2} { \ddot{u}^2 \over  \dot{u}^3 }
+ { \ddot{u} \over  \dot{u}^2 }
\right)
- x { \ddot{u} \over \dot{u} }
- 1 , \\
\tilde{I}_{3b}
=
u^2 \left(
{ x \over 2} { \ddot{u}^2 \over  \dot{u}^3 }
+ { \ddot{u} \over  \dot{u}^2 }
\right)
- 2 u \left(
 x { \ddot{u} \over \dot{u} }
+  1
\right)
+ 2 x \dot{u}  ,  \\
\tilde{I}_{4b}
\equiv   0 ,
\qquad
\tilde{I}_{5b} \equiv  1 ,
\qquad
\tilde{I}_{6b} \equiv  0 .
\end{multline}
For $ v _c (x) = x^2  $ we find
\begin{multline}    \label{firstint1c}
\tilde{I} _{1c}
= { x^2  \over 2} { \ddot{u}^2 \over  \dot{u}^3 }
+ 2x { \ddot{u} \over  \dot{u}^2 }
+ { 2  \over  \dot{u} }   ,
\qquad
\tilde{I}_{2c}
= u \left(
{ x^2  \over 2} { \ddot{u}^2 \over  \dot{u}^3 }
+ 2x { \ddot{u} \over  \dot{u}^2 }
+ { 2  \over  \dot{u} }
\right)
- x ^2 { \ddot{u} \over \dot{u} }
- 2 x   ,  \\
\tilde{I}_{3c}
= u^2  \left(
{ x^2  \over 2} { \ddot{u}^2 \over  \dot{u}^3 }
+ 2x { \ddot{u} \over  \dot{u}^2 }
+ { 2  \over  \dot{u} }
\right)
- 2 u \left(
 x ^2 { \ddot{u} \over \dot{u} }
+  2 x \right)
+ 2 x^2 \dot{u}   ,   \\
\tilde{I}_{4c}
\equiv   -2 ,
\qquad
\tilde{I}_{5c} \equiv  0 ,
\qquad
\tilde{I}_{6c} \equiv  0 .
\end{multline}

The nontrivial first integrals given in~(\ref{firstint1b}),(\ref{firstint1c})
together with nontrivial first integrals given in~(\ref{common1})
satisfy the relations
$$
\tilde{I}_{1a} \tilde{I}_{3a} - \tilde{I}_{2a}  ^2 = 0   ,
\qquad
\tilde{I}_{1b} \tilde{I}_{3b} - \tilde{I}_{2b}  ^2 = -1  ,
\qquad
\tilde{I}_{1c} \tilde{I}_{3c} - \tilde{I}_{2c}  ^2 = 0   ,
$$
$$
\tilde{I}_{1a} \tilde{I}_{1c} - \tilde{I}_{1b}  ^2 = 0   ,
\qquad
\tilde{I}_{2a} \tilde{I}_{2c} - \tilde{I}_{2b}  ^2 = - 1   ,
\qquad
\tilde{I}_{3a} \tilde{I}_{3c} - \tilde{I}_{3b}  ^2 = 0   .
$$

\bigskip

\noindent {\bf Integration of the ODE}

\medskip

We chose three first integrals  $\tilde{I}_{1a}$, $\tilde{I}_{2a}$ and $\tilde{I}_{1b}$
and compute the Jacobian
$$
J =  \mbox{det} \left(
{ \partial ( \tilde{I}_{1a}, \tilde{I}_{2a}, \tilde{I}_{1b}  )\over
 \partial  (u, \dot{u} , \ddot{u} ) } \right)
 =  - { 1 \over 4 }  { \ddot{u} ^2 \over \dot{u} ^9 }   .
$$

\begin{enumerate}

\item

For $J \neq 0$ we can set these first integrals equal to constants
$$
\tilde{I}_{1a}
=   { 1\over 2}  { \ddot{u}^2 \over \dot{u}^3 } = A  ,
\qquad
\tilde{I}_{2a}
= { u \over 2} { \ddot{u}^2 \over \dot{u}^3 }
- { \ddot{u} \over \dot{u} }   = B   ,
\qquad
\tilde{I} _{1b}
= { x \over 2} { \ddot{u}^2 \over  \dot{u}^3 }
+ { \ddot{u} \over  \dot{u}^2 }  = C ,
$$
where $ A  \neq  0$, $B$ and $C$ are constants.
We rewrite the second and third equations as
$$
  A u    - { \ddot{u} \over \dot{u} }   = B   ,
\qquad
  A x   + { \ddot{u} \over  \dot{u}^2 }  = C
$$
and exclude the derivatives
$$
{ \ddot{u} \over \dot{u} } { \ddot{u} \over  \dot{u}^2 }
= ( A u    - B  ) ( C - A x ) = A .
$$
From this equation we can obtain the solution
which we present as
\begin{equation}
u (x) = { 1 \over C_1 x + C_2 } +  C_3 ,
\end{equation}
where $C_1 \neq 0 $, $C_2$ and $C_3$  are constants.
It is the generic (three-parameter) solution of the ODE.

\item

Considering $J = 0$, we solve
$$
 \ddot{u} =  0 ,   \qquad  \dot{u} \neq 0
$$
and obtain
\begin{equation}
u(x) = C_1 x + C_2 , \qquad C_1 \neq 0 .
\end{equation}
Direct verification shows that it is a solution of the ODE~(\ref{third}).
It depends only on two parameters and is thus a degenerate solution.

\end{enumerate}

\bigskip

\noindent {\bf Case:  $ M < 0 $}

\medskip

The specific solutions of the adjoint equation~(\ref{adjoint0}) are~(\ref{328a}) and
\begin{equation*}
v _b (x) = \cosh( 2 \omega x)
\qquad
\mbox{and}
\qquad
v _c (x) = \sinh( 2 \omega x) ,
\qquad
\omega  = \sqrt{- {M\over 2}} .
\end{equation*}
For these solutions $ v _b$ and $ v _c$ we find first integrals
$$
\tilde{I} _{1b}
= \cosh( 2 \omega x)
\left( {1 \over 2}    { \ddot{u}^2 \over  \dot{u}^3 }  - {M \over \dot{u} } \right)
+ 2 \omega \sinh( 2 \omega x)
\left( {  \ddot{u}   \over \dot{u}^2 } \right) ,
$$
\begin{equation}   \label{firstint2b}
\tilde{I}_{2b}
= \cosh( 2 \omega x) \left(
u \left(
{ 1\over 2} {  \ddot{u}^2 \over \dot{u}^3 }
-  { M \over \dot{u} } \right)
- { \ddot{u} \over \dot{u} }
\right)
+ 2 \omega \sinh( 2 \omega x)
\left( u {  \ddot{u}   \over \dot{u}^2 } - 1  \right) ,
\end{equation}
$$
\tilde{I}_{3b}
= \cosh( 2 \omega x) \left(
u^2 \left(
{ 1\over 2}  { \ddot{u}^2 \over  \dot{u}^3 }
- { M   \over \dot{u} } \right)
- 2 u {  \ddot{u} \over \dot{u} }
+ 2 \dot{u}   \right)
+ 2 \omega \sinh( 2 \omega x)
\left( u^2 {  \ddot{u}   \over \dot{u}^2 } -  2 u   \right) ,
$$
$$
\tilde{I}_{4b}
\equiv  0
$$
and
$$
\tilde{I} _{1c}
= \sinh( 2 \omega x)
\left( {1 \over 2} { \ddot{u}^2 \over \dot{u}^3 }  - {M \over \dot{u} } \right)
+  2 \omega \cosh( 2 \omega x)
\left( {  \ddot{u}   \over \dot{u}^2 } \right)  ,
$$
\begin{equation}   \label{firstint2c}
\tilde{I}_{2c}
= \sinh( 2 \omega x) \left(
u \left(
{ 1\over 2} { \ddot{u}^2 \over \dot{u}^3 }
-   { M  \over \dot{u} } \right)
- { \ddot{u} \over \dot{u} }
\right)
+  2 \omega \cosh( 2 \omega x)
\left( u {   \ddot{u}   \over \dot{u}^2 } -1  \right)  ,
\end{equation}
$$
\tilde{I}_{3c}
= \sinh( 2 \omega x) \left(
u^2 \left(
{ 1\over 2}  {  \ddot{u}^2 \over  \dot{u}^3 }
-  { M  \over \dot{u} } \right)
- 2 u {  \ddot{u} \over \dot{u} }
+ 2 \dot{u}   \right)
+ 2 \omega \cosh( 2 \omega x)
\left(   u^2  { \ddot{u}   \over \dot{u}^2 } -  2u   \right) ,
$$
$$
\tilde{I}_{4c}
\equiv  0  ,
$$
respectively.

Nontrivial first integral of this case,
which are given in~(\ref{common1}), (\ref{firstint2b}) and~(\ref{firstint2c}),
obey the relations
$$
\tilde{I}_{1a} \tilde{I}_{3a} - \tilde{I}_{2a}  ^2 = 2 M  ,
\qquad
\tilde{I}_{1b} \tilde{I}_{3b} - \tilde{I}_{2b}  ^2 = - 2 M  ,
\qquad
\tilde{I}_{1c} \tilde{I}_{3c} - \tilde{I}_{2c}  ^2 = 2 M  ,
$$
$$
\tilde{I}_{1a}  ^2 + \tilde{I}_{1c} ^2  = \tilde{I}_{1b} ^2 ,
\qquad
\tilde{I}_{2a}  ^2 + \tilde{I}_{2c} ^2  = \tilde{I}_{2b} ^2 - 2 M  ,
\qquad
\tilde{I}_{3a}  ^2 + \tilde{I}_{3c} ^2  = \tilde{I}_{3b} ^2  .
$$

\bigskip

\noindent {\bf Integration of the ODE}

\medskip

We select  $\tilde{I}_{1a}$, $\tilde{I}_{2a}$ and $\tilde{I}_{1b}$
and compute the Jacobian
$$
J = \mbox{det} \left(
{ \partial  ( \tilde{I}_{1a}, \tilde{I}_{2a}, \tilde{I}_{1b} )  \over
 \partial  ( u, \dot{u} , \ddot{u} )  } \right)
$$
$$
= - { \omega \over 2 }
{ \ddot{u} ^2  - 4 \omega ^2  \dot{u} ^2   \over   \dot{u} ^9 }
\left(
\sinh ( 2 \omega x )  \ddot{u} ^2
+ 4  \omega  \cosh ( 2 \omega x ) \dot{u}   \ddot{u}
+ 4  \omega ^2 \sinh ( 2 \omega x ) \dot{u} ^2
\right)  .
$$

\begin{enumerate}

\item

In the case  $J \neq 0 $ we can set these first integrals equal to constants
and obtain solutions
\begin{equation}   \label{sol2part1}
u (x) =  C_1 \tanh ( \omega x + C_2 ) + C_3
\quad
\mbox{and}
\quad
u (x) =  C_1 \coth ( \omega x + C_2 ) + C_3 ,
\end{equation}
where $C_1 \neq 0 $, $C_2 \neq 0 $ and $C_3$  are constants.

\item

The case  $J = 0 $ gets split into two subcases.

\begin{enumerate}

\item

The system
$$
 \ddot{u} ^2  - 4 \omega ^2  \dot{u} ^2   = 0 , \qquad   \dot{u}  \neq 0
$$
leads to
$$
\ddot{u}   = \pm 2 \omega  \dot{u}  ,
\qquad
\dot{u} \neq 0 .
$$
We obtain two degenerate solutions of the ODE
\begin{equation}    \label{sol2part3}
u(x) = C_1 e^{ \pm 2 \omega x} + C_2 , \qquad C_1 \neq 0 .
\end{equation}

\item

The  system
$$
\sinh ( 2 \omega x )  \ddot{u} ^2
+ 4  \omega  \cosh ( 2 \omega x ) \dot{u}   \ddot{u}
+ 4  \omega ^2  \sinh ( 2 \omega x ) \dot{u} ^2
= 0 ,
\qquad
\dot{u}  \neq 0
$$
can be solved  as quadratic for $ \ddot{u}$:
$$
\ddot{u} = - 2 \omega   \coth ( \omega x ) \dot{u}
\qquad
\mbox{or}
\qquad
\ddot{u} = - 2 \omega   \tanh  ( \omega x ) \dot{u}   ,
\qquad
\dot{u}  \neq 0  ,
$$
which can be solved as
\begin{equation}  \label{sol2part2}
u (x) =  C_1 \tanh ( \omega x ) + C_2
\quad
\mbox{or}
\quad
u (x) =  C_1 \coth ( \omega x ) + C_2  ,
\qquad
C_1 \neq 0 .
\end{equation}

\end{enumerate}

\end{enumerate}

Finally solutions~(\ref{sol2part1}), (\ref{sol2part3}) and~(\ref{sol2part2})
can be written together as given in~(\ref{sol2}).

\eject


\bigskip

\noindent {\bf Case:  $ M >  0 $}

\medskip

The specific solutions of the adjoint equation are~(\ref{328a}) and
\begin{equation*}
v _b (x) = \cos( 2 \omega x)
\qquad
\mbox{and}
\qquad
v _c (x) = \sin( 2 \omega x) ,
\qquad
\omega  = \sqrt{M \over 2} .
\end{equation*}
For $v _b$ we compute the first integrals
$$
\tilde{I} _{1b}
= \cos ( 2 \omega x)
\left( {1 \over 2} { \ddot{u}^2 \over  \dot{u}^3 }  - {M \over \dot{u} } \right)
- 2 \omega \sin ( 2 \omega x)
\left( {  \ddot{u}   \over \dot{u}^2 } \right) ,
$$
\begin{equation}   \label{firstint3b}
\tilde{I}_{2b}
= \cos ( 2 \omega x) \left(
u \left(
{ 1\over 2} { \ddot{u}^2 \over \dot{u}^3 }
-   { M  \over \dot{u} } \right)
- { \ddot{u} \over \dot{u} } \right)
- 2 \omega \sin ( 2 \omega x)
\left( u {   \ddot{u}   \over \dot{u}^2 } -1  \right) ,
\end{equation}
$$
\tilde{I}_{3b}
= \cos( 2 \omega x) \left(
u^2 \left(
{ 1\over 2}  {  \ddot{u}^2 \over  \dot{u}^3 }
-  { M   \over \dot{u} } \right)
- 2 u {   \ddot{u} \over \dot{u} }
+ 2 \dot{u}   \right)
- 2 \omega \sin ( 2 \omega x)
\left(   u^2  { \ddot{u}   \over \dot{u}^2 } -  2 u   \right) ,
$$
$$
\tilde{I}_{4b}
\equiv  0  .
$$
For $v _c$ we get
$$
\tilde{I} _{1c}
= \sin( 2 \omega x)
\left( {1 \over 2} { \ddot{u}^2 \over  \dot{u}^3 }  - {M \over \dot{u} } \right)
+  2 \omega \cos( 2 \omega x)
\left( {  \ddot{u}   \over \dot{u}^2 } \right) ,
$$
\begin{equation}   \label{firstint3c}
\tilde{I}_{2c}
= \sin ( 2 \omega x) \left(
u \left(
{ 1\over 2} {  \ddot{u}^2 \over \dot{u}^3 }
- { M  \over \dot{u} } \right)
- { \ddot{u} \over \dot{u} }
 \right)
+ 2 \omega \cos ( 2 \omega x)
\left( u {  \ddot{u}   \over \dot{u}^2 } - 1  \right) ,
\end{equation}
$$
\tilde{I}_{3c}
= \sin ( 2 \omega x) \left(
u^2 \left(
{ 1\over 2}  { \ddot{u}^2 \over  \dot{u}^3 }
- { M   \over \dot{u} } \right)
- 2 u {  \ddot{u} \over \dot{u} }
+ 2 \dot{u}   \right)
+ 2 \omega \cos ( 2 \omega x)
\left(  u^2 {  \ddot{u}   \over \dot{u}^2 } -  2 u   \right) ,
$$
$$
\tilde{I}_{4c}
\equiv  0  .
$$

The nontrivial first integrals~(\ref{common1}),(\ref{firstint3b}),(\ref{firstint3c})
satisfy the relations
$$
\tilde{I}_{1a}  \tilde{I}_{3a} - \tilde{I}_{2a}  ^2 = 2 M  ,
\qquad
\tilde{I}_{1b}  \tilde{I}_{3b} - \tilde{I}_{2b}  ^2 = - 2 M  ,
\qquad
\tilde{I}_{1c}  \tilde{I}_{3c} - \tilde{I}_{2c}  ^2 = - 2 M  ,
$$
$$
\tilde{I}_{1b}  ^2 + \tilde{I}_{1c} ^2  = \tilde{I}_{1a} ^2 ,
\qquad
\tilde{I}_{2b}  ^2 + \tilde{I}_{2c} ^2  = \tilde{I}_{2a} ^2 + 2 M  ,
\qquad
\tilde{I}_{3b}  ^2 + \tilde{I}_{3c} ^2  = \tilde{I}_{3a} ^2 .
$$

\bigskip

\noindent {\bf Integration of the ODE}

\medskip

Similarly to the previous cases we chose   $\tilde{I}_{1a}$, $\tilde{I}_{2a}$ and $\tilde{I}_{1b}$
as three first integrals.
The Jacobian is
$$
J = \mbox{det} \left(
{ \partial  ( \tilde{I}_{1a}, \tilde{I}_{2a}, \tilde{I}_{1b} )  \over
 \partial  (u, \dot{u} , \ddot{u} )  } \right)
$$
$$
 =  { \omega \over 2 }
{ \ddot{u} ^2  + 4 \omega ^2  \dot{u} ^2   \over   \dot{u} ^9 }
\left(
\sin ( 2 \omega x )  \ddot{u} ^2
+ 4  \omega  \cos ( 2 \omega x )  \dot{u}   \ddot{u}
- 4  \omega ^2  \sin ( 2 \omega x ) \dot{u} ^2
\right)  .
$$

\begin{enumerate}

\item

For $J \neq 0 $ we obtain the solutions
\begin{equation}    \label{sol3part1}
u (x) = C_1 \tan ( \omega x + C_2 ) + C_3 ,
\end{equation}
where $C_1 \neq 0 $, $C_2 \neq { \pi  \over 2} n $, $ n   \in \mathbb{Z} $
and $C_3$  are constants.

\item

The equality  $J  = 0 $ can happen in two cases

\begin{enumerate}

\item

First system
$$
\ddot{u} ^2  + 4 \omega ^2  \dot{u} ^2 = 0 ,
\qquad
\dot{u} \neq 0
$$
has no solutions.

\item

Second system
$$
\sin ( 2 \omega x )  \ddot{u} ^2
+ 4  \omega  \cos ( 2 \omega x )  \dot{u}   \ddot{u}
- 4  \omega ^2  \sin ( 2 \omega x )  \dot{u} ^2
= 0 ,
\qquad
\dot{u} \neq 0
$$
can be solved for $ \ddot{u}$:
leads to
$$
\ddot{u} = 2 \omega  \tan  ( \omega x )  \dot{u}
\qquad
\mbox{or}
\qquad
\ddot{u} = - 2 \omega  \cot  ( \omega x )  \dot{u}  ,
\qquad
\dot{u} \neq 0 .
$$
These equations are solved as
\begin{equation}    \label{sol3part2}
u (x) = C_1 \tan ( \omega x ) + C_2
\qquad
\mbox{or}
\qquad
u (x) = C_1 \cot ( \omega x ) + C_2 ,
\end{equation}
where $C_1 \neq 0 $ and $C_2$  are integration constants.

\end{enumerate}

\end{enumerate}

Finally, we unite solutions~(\ref{sol3part1}) and~(\ref{sol3part2})
into the generic solution of the ODE given in~(\ref{sol3}).


\begin{remark}
It is possible to use a different Ansatz for solutions of the
adjoint equation.

\bigskip

\noindent { \bf Ansatz~2.} If we look for solutions of the form $ v
=  v (x,u,\dot{u}) $, we find
$$
v (x,u,\dot{u}) = { A ( x,u) \over \sqrt{| \dot{u} |} }
+ { B ( u) \over \dot{u} }
+  C(x)  ,
$$
where
$$
A(x,u) = \alpha (x) u + \beta (x) , \qquad
 \ddot{\alpha}  + {M \over 2}  \alpha = 0  ,
\qquad
 \ddot{\beta}  + {M \over 2}  \beta  = 0  ,
$$
$$
B(u) = \gamma _2 u^2 + \gamma _1 u + \gamma _0 ,
\qquad
\dddot{C} + 2 M \dot{C} = 0  .
$$
Here we have to consider cases  $M=0$, $M<0$ and $M>0$ separately.
We obtain the following independent solutions for $ \alpha (x)$ and $ \beta (x)$:
$$
\begin{array}{llll}
 M = 0 : &  {\alpha}_1 = 1  ,  & {\alpha}_2 = x, &  \\
  &   &  &  \\
         &   {\beta}_1 = 1  ,  & {\beta}_2 = x;   &       \\
  &   &  &  \\
 M < 0 : & {\alpha}_1 = \cosh ( \omega  x )  , & {\alpha}_2 = \sinh ( \omega  x )  , &  \\
  &   &  &  \\
         & {\beta}_1 = \cosh ( \omega  x )  , & {\beta}_2 =\sinh ( \omega  x )  , &
     \omega = \sqrt{ - { M / 2}} ;  \\
  &   &  &  \\
 M > 0 : & {\alpha}_1 = \cos ( \omega  x )  , & {\alpha}_2 = \sin ( \omega  x )  , & \\
  &   &  &  \\
         &   {\beta}_1 = \cos ( \omega  x )  , & {\beta}_2 = \sin ( \omega  x )  , &
    \omega = \sqrt{  { M / 2}} . \\
\end{array}
$$
Solutions for $C(x) $ are the same as solutions for $v(x) $ given
in~(\ref{indep1}). In each case we will obtain 10 independent
solutions of the adjoint equation.
The solutions of the original ODE~(\ref{third}) are of course the same
as obtained with Ansatz~1.
\end{remark}

\section{The direct method and integrating factors}

\label{integrating}

In this section we compare the direct method~\cite{[Blu]}
with the method presented in the previous section.
We consider a scalar ODE
\begin{equation}
 F ( x, u, \dot{u}, \ddot{u},  ... , u^{(n)} ) = 0
\end{equation}
and assume that this ODE is solved with respect to the highest derivative  $u^{(n)}$:
\begin{equation}
F =   u^{(n)} -  f ( x, u, \dot{u}, \ddot{u},  ... , u^{(n-1)} ) = 0 .
\end{equation}

We are interested in first integrals
\begin{equation}
I =  I (x,u,\dot{u},..., u^{(n-1)} )
\end{equation}
of this ODE such that
\begin{equation}  \label{forintegral}
D ( I ) = \Lambda  F ,
\end{equation}
 holds identically in the whole space for some nonsingular function
\begin{equation}
\Lambda = \Lambda(x,u,\dot{u},..., u^{(n-1)} ) ,
\end{equation}
called an {\it integrating factor}.
Since the left hand side of relation~(\ref{forintegral}) is a total derivative
it is annihilated by the action of the variational operator.
Therefore we obtain the equation
\begin{equation}  \label{factor}
{\delta {} \over \delta u } (\Lambda F) = 0 .
\end{equation}

\begin{remark}   \label{linearity}
The assumption that the ODE  is solved with respect to the highest
derivative $u^{(n)}$ allows us to restrict the form of the integrating
factor to $\Lambda = \Lambda(x,u,\dot{u},..., u^{(n-1)} )$
(otherwise $\Lambda $ could depend on the highest derivative). It
provides us with the possibility of splitting equation~(\ref{factor})
into  $1 + [ n/2 ] $  equations~\cite{[Blu]}.
\end{remark}

Let us provide comparison with the adjoint equation
\begin{equation}  \label{adjn}
\left.
{\delta {}\over \delta u } (v F)
\right|_{F=0} = 0 ,
\qquad
v = \varphi ( x, u, \dot{u}, ..., u^{(n-1)} )  .
\end{equation}
If  we consider $\Lambda$ and $ \varphi$ which depend on the same variables,
we obtain the following result.

\begin{proposition}
An integrating factor is always a   solution of adjoint equation
independently of whether a symmetry of the underlying equation exists.
The inverse statement is not true.
\end{proposition}


Using integrating factors,
we have to solve the equation~(\ref{forintegral})
in order to find a first integral $  I $.
Explicit line integral formulas which provide first integrals were given in~\cite{[Blu]}.
The approach based on the solution of the adjoint equation yields
the first integral by formula~(\ref{firstint1}),
which does not require any integration.




It was observed on the examples of the previous section that
the pair consisting  of a symmetry $X$ and a solution of the adjoint equation $v$
can generate a trivial first integral.
To the contrary, an integrating factor $\Lambda$ provides a non-trivial first integral.
On the other hand,
it will be observed on the example given below that the approach based on Theorem~\ref{main}
has the  advantage that we can use a simpler Ansatz for  $ \varphi$ then for  $\Lambda$.

\bigskip

\noindent {\bf Example: application of the direct method}

\medskip

We demonstrate the direct method using the example of ODE~(\ref{third}).
We multiply the equation by the unknown factor $ \Lambda \left(  x, u,
\dot{u}, \ddot{u} \right)$ and apply a variational operator~(\ref{euler}):
\begin{equation} \label{direct_1}
\frac{\delta}{\delta u} \left(
        \Lambda \left(  x, u, \dot{u}, \ddot{u} \right) \left(
                \frac{\dddot{u}}{\dot{u}} - \frac{3}{2} \left(  \frac{\ddot{u}}{\dot{u}} \right)^2 - M
            \right)
    \right) = 0 .
\end{equation}

To simplify the problem we are looking for an integrating factor of the simplest form.
It can be shown that there are no integrating factors of the form  $  \Lambda (  x, u )$.
Therefore we chose the form
\begin{equation}
\Lambda = \Lambda (  x, u, \dot{u} ).
\end{equation}
Equation~(\ref{direct_1}) appears as
\begin{equation}
- \left(
    \frac{2}{\dot{u}} \, \frac{\partial  \Lambda \left(  x, u, \dot{u} \right)}{\partial \dot{u} }
    + \frac{ \Lambda \left(  x, u, \dot{u} \right)}{\dot{u}^2}
\right) \,  u^{(4)} + ... ,
\end{equation}
where the rest of equation does not contain $u^{(4)} $.
Thus we get the equation
\begin{equation}
    \frac{2}{\dot{u}} \, \frac{\partial  \Lambda \left(  x, u, \dot{u} \right)}{\partial \dot{u}}
    + \frac{ \Lambda \left(  x, u, \dot{u} \right)}{\dot{u}^2} = 0 .
\end{equation}
Integration for variable $\dot{u}$ yields
\begin{equation}   \label{formL}
\Lambda = \frac{f(x,u)}{\sqrt{|\dot{u}|}},
\end{equation}
where $f(x,u)$ is some function.

Substitution of  $\Lambda$  into~(\ref{direct_1}) yields
\begin{multline}   \label{subs}
\frac{3}{4  \sqrt{|\dot{u}|}}
\left(
    \frac{2  f_{xx}  + M f } {\dot{u}^2} - 2  f_{uu}
\right)  \ddot{u}
- |\dot{u}|^{\frac{3}{2}}   f_{uuu}
- 3  \mbox{sgn} (\dot{u}) \sqrt{|\dot{u}|}   f _{xuu}  \\
 - \frac{3}{2 \, \sqrt{|\dot{u}|}}
 ( 2  f_{xxu}   +   M f_u  )
- \frac{1}{2}   \mbox{sgn} (\dot{u})  |\dot{u}|^{-\frac{3}{2}}
 ( 2  f_{xxx}  +     M  f_x )
  = 0   ,
\end{multline}
where
$$
\mbox{sgn} (x)
= \left\{
\begin{array}{rl}
1 , &  x > 0 ; \\
0 , &  x = 0 ; \\
-1 , & x < 0 . \\
\end{array}
\right.
$$
Equating to zero the coefficients of different powers of  $\ddot{u}$ and
then those of different powers of $\dot{u}$,
we obtain the system
\begin{equation}
f_{uu} = 0 ,
\qquad
f_{xx}  + {M \over 2} f  = 0 .
\end{equation}
This system has the following solutions for different cases of $M$:
\begin{equation}
M = 0:
\qquad
f(x,u) = ( C_1 x + C_2 ) u   + C_3 x + C_4 ;
\end{equation}
\begin{multline}
M < 0:
\qquad
f(x,u) = (  C_1 e^{ \omega x }  + C_2 e^{ - \omega x }   ) u   \\
+  C_3 e^{ \omega x }  + C_4 e^{ - \omega x }  ,
\qquad
\omega = \sqrt{- M / 2} ;
\end{multline}
\begin{multline}
M > 0:
\qquad
f(x,u) = (C_1 \cos ( \omega x )  + C_2 \sin ( \omega x ) ) u  +   \\
+   C_3 \cos ( \omega x )  + C_4 \sin ( \omega x ) ,
\qquad
\omega = \sqrt{ M / 2 } .
\end{multline}

Thus we can present a number of independent solutions $\Lambda$
and find the corresponding integrals $ I (x, u, \dot{u}, \ddot{u})$
using the relation
\begin{equation}    \label{forfirst}
D ( I )
=  {\partial  I \over \partial  x }
+   \dot{u} {\partial  I  \over \partial  u }
+   \ddot{u} {\partial  I  \over \partial  \dot{u}  }
+   \dddot{u} {\partial  I  \over \partial  \ddot{u}  }
= \Lambda F  .
\end{equation}
This equation is solved in the same manner as~(\ref{subs}),
starting from the highest derivative $\dddot{u} $.

\bigskip

\noindent {\bf Case: $M = 0$}

\medskip

There are four integrating factors
$$
\Lambda _1 = { 1 \over \sqrt{|\dot{u}|}  }  ,
\qquad
\Lambda _2 = { x \over \sqrt{|\dot{u}|}  }  ,
\qquad
\Lambda _3 = { u \over \sqrt{|\dot{u}|}  }  ,
\qquad
\Lambda _4 = { x u \over \sqrt{|\dot{u}|}  }  ,
$$
which provide us with the corresponding first integrals
\begin{multline}
I_1 = { 1 \over \sqrt{|\dot{u}|}  }  { \ddot{u}   \over \dot{u}  }   ,
\qquad
I_2 = { 1 \over \sqrt{|\dot{u}|}  }
\left(
x { \ddot{u}   \over \dot{u}  }   + 2
\right) , \\
I_3 = { 1 \over \sqrt{|\dot{u}|}  }
\left(
u { \ddot{u}   \over \dot{u}  }   - 2 \dot{u}
\right) ,
\qquad
I_4 = { 1 \over \sqrt{|\dot{u}|}  }
\left(
x u { \ddot{u}   \over \dot{u}  }   - 2 x \dot{u} + 2 u
\right) ,
\end{multline}
which obey the relation
$$
I_1 I_4 -  I_2 I_3  =  4 \mbox{sgn} ( \dot{u} )     .
$$

\bigskip

\noindent {\bf Integration of the ODE}

\medskip

We select first integrals $ I_1$, $ I_2$ and $ I_3$
and compute the Jacobian
\begin{equation*}
J = \mbox{det}
\left(
\frac{\partial( I_1, I_2 , I_3 )}{\partial(u, \dot{u}, \ddot{u})}
\right)=   \mbox{sgn} (\dot{u}) \frac{\ddot{u}}{ |\dot{u}|^{\frac{9}{2}}}   .
\end{equation*}

\begin{enumerate}

\item

In the case $J \neq 0$ we can use the first integrals.
Setting them to be equal to constants
$$
I_1 = { 1 \over \sqrt{|\dot{u}|}  }  { \ddot{u}   \over \dot{u}  } = \mbox{const}   ,
\qquad
I_2 = { 1 \over \sqrt{|\dot{u}|}  }
\left(
x { \ddot{u}   \over \dot{u}  }   + 2
\right)  = \mbox{const} ,
$$
$$
I_3 = { 1 \over \sqrt{|\dot{u}|}  }
\left(
u { \ddot{u}   \over \dot{u}  }   - 2 \dot{u}
\right)  = \mbox{const}  ,
$$
we can find the the generic solution
\begin{equation}
u(x) = { 1 \over  C_1 x + C_2} + C_3 ,
\end{equation}
where $C_1 \neq 0$, $C_2$, and $C_3$ are constants.

\item

The case $J = 0 $ leads to the system
$$
\ddot{u} = 0 ,
\qquad
\dot{u} \neq 0  .
$$
Thus, we obtain the special solution using the ODE
\begin{equation}
u(x) = C_1 x + C_2,
\qquad
C_1 \neq 0 .
\end{equation}

\end{enumerate}

\bigskip

\noindent {\bf Case:  $M < 0$}

\medskip

There are four independent integrating factors
$$
\Lambda _1 = { e^{ \omega x }  \over \sqrt{|\dot{u}|}  }  ,
\qquad
\Lambda _2 = { e^{ - \omega x }  \over \sqrt{|\dot{u}|}  }  ,
\qquad
\Lambda _3 = { e^{ \omega x }  u \over \sqrt{|\dot{u}|}  }  ,
\qquad
\Lambda _4 = { e^{ - \omega x }  u \over \sqrt{|\dot{u}|}  }  .
$$
They generate the  first integrals
\begin{multline}
I_1 = { e^{ \omega x }  \over \sqrt{|\dot{u}|}  }
\left(
 { \ddot{u}   \over \dot{u}  }   + 2  \omega
\right) ,
\qquad
I_2 = { e^{ - \omega x }  \over \sqrt{|\dot{u}|}  }
\left(
 { \ddot{u}   \over \dot{u}  }   - 2  \omega
\right) , \\
I_3 = { e^{ \omega x }  \over \sqrt{|\dot{u}|}  }
\left(
u { \ddot{u}   \over \dot{u}  }  - 2 \dot{u}  + 2  \omega u
\right) ,
\qquad
I_4 = { e^{ - \omega x }  \over \sqrt{|\dot{u}|}  }
\left(
u { \ddot{u}   \over \dot{u}  }   - 2 \dot{u} - 2  \omega u
\right) .
\end{multline}
These four first integrals obey the relation
$$
I_1 I_4  - I_2 I_3 =  - 8  \omega  \mbox{sgn} ( \dot{u} )  .
$$

\bigskip

\noindent {\bf Integration of the ODE }

\medskip

We chose the three first integrals $I_1$, $I_2$ and $I_3$,
compute the Jacobian
\begin{equation*}
\mbox{det}
\left(
\frac{\partial( I_1, I_2 , I_3 )}{\partial(u, \dot{u}, \ddot{u})}
\right)
=
- 2 \omega e^{ \omega x }  \mbox{sgn} (\dot{u})
{ \ddot{u} + 2  \omega \dot{u} \over   |\dot{u}|^{\frac{9}{2}}   }
\end{equation*}
and consider two cases.

\begin{enumerate}

\item

In the case $J\neq 0$ we can use the values of the first integrals
to obtain the generic solutions of the ODE
\begin{equation}
u (x) =  C_1 \tanh ( \omega x + C_2 ) + C_3
\qquad
\mbox{and}
\qquad
u (x) =  C_1 \coth ( \omega x + C_2 ) + C_3 ,
\end{equation}
where $C_1 \neq 0 $, $C_2$ and $C_3$ are constants,
as well as one special solution
\begin{equation}
u (x) = C_1 e^{ 2 \omega  x }  + C_2 ,
\qquad
C_1 \neq 0 .
\end{equation}

\item

If $J = 0 $, we obtain the system
$$
\ddot{u} + 2  \omega \dot{u}  = 0 ,
\qquad
\dot{u} \neq 0  .
$$
and find the special solution
\begin{equation}
u(x) = C_1  e^{ -2 \omega x }   + C_2,
\qquad
C_1 \neq 0 .
\end{equation}

\end{enumerate}

\bigskip

\noindent {\bf Case: $M > 0$}

\medskip

As in the previous case there are four independent integrating factors
$$
\Lambda _1 = { \cos  (\omega x )  \over \sqrt{|\dot{u}|}  }  ,
\qquad
\Lambda _2 = { \sin  (\omega x )  \over \sqrt{|\dot{u}|}  }  ,
\qquad
\Lambda _3 = { \cos  (\omega x )  u   \over \sqrt{|\dot{u}|}  }  ,
\qquad
\Lambda _4 = { \sin  (\omega x )  u    \over \sqrt{|\dot{u}|}  }  ,
$$
which let us find the corresponding first integrals
\begin{multline}
I_1 = { 1 \over \sqrt{|\dot{u}|}  }
\left(
\cos  (\omega x )  { \ddot{u}   \over \dot{u}  }   - 2  \omega  \sin  (\omega x )
\right) ,
\qquad
I_2 = { 1 \over \sqrt{|\dot{u}|}  }
\left(
\sin  (\omega x )   { \ddot{u}   \over \dot{u}  }   + 2  \omega \cos  (\omega x )
\right) ,
\\
I_3 = { 1 \over \sqrt{|\dot{u}|}  }
\left(
\cos  (\omega x )  \left( u {  \ddot{u}   \over \dot{u}  }  - 2 \dot{u} \right)
- 2  \omega  \sin  (\omega x ) u
\right) ,
\\
I_4 = { 1 \over \sqrt{|\dot{u}|}  }
\left(
\sin  (\omega x )  \left(  u {  \ddot{u}   \over \dot{u}  } - 2 \dot{u} \right)
+ 2  \omega \cos  (\omega x )  u
\right) .
\end{multline}
The first integrals satisfy the relation
$$
I_1 I_4  - I_2 I_3   =  4  \omega \mbox{sgn} (\dot{u}) .
$$

\bigskip

\noindent {\bf Integration of the ODE}

\medskip

We find the Jacobian
\begin{equation*}
\mbox{det}
\left(
\frac{\partial( I_1, I_2 , I_3 )}{\partial(u, \dot{u}, \ddot{u})}
\right)
=  \omega  \mbox{sgn} (\dot{u})
{  \cos  (\omega x )  \ddot{u} - 2  \omega    \sin  (\omega x ) \dot{u}  \over   |\dot{u}|^{\frac{9}{2}}   }
\end{equation*}
and consider two cases.

\begin{enumerate}

\item

For $J \neq 0$ we obtain the solution in the form
\begin{equation}
u (x) = C_1 \tan ( \omega x + C_2 ) + C_3 ,
\end{equation}
where  $C_1 \neq 0$, $C_2 \neq  \pi n $, $ n   \in \mathbb{Z} $ and  $C_3$
are constants.

\item

The determinant of the Jacobian equals to zero if
$$
 \cos  (\omega x )  \ddot{u} - 2  \omega    \sin  (\omega x ) \dot{u}  = 0 ,
\qquad
\dot{u} \neq 0  .
$$
Thus, we obtain the special solution
\begin{equation}
u(x) = C_1  \tan ( \omega x )  + C_2,
\qquad
C_1 \neq 0 .
\end{equation}

\end{enumerate}

\bigskip

Let us sum up the comparison of the two methods for the example~(\ref{third}).

\begin{enumerate}

\item

Both methods allow to find the general solution;

\item

Integration factors as opposed to the adjoint equation method
always provide non-trivial first integrals;

\item

We had to use a more complicated Anzats for $\Lambda$ than for $v$.
Searching for integrating factors, we needed $\Lambda =  \Lambda  ( x, u, \dot{u} ) $.
It was sufficient to consider the simple Ansatz  $v  =  v  ( x ) $
for the solution of the adjoint equation;

\item

Using $\Lambda$ in the direct method, we have to integrate for obtaining first integrals.
Using $v$ and symmetry operator $X$ in the adjoint equation method,
we apply a formula which does not require integration.
This advantage may be crucial in the case of discrete equations.

\end{enumerate}

\section{Adjoint equation method for mappings}

\label{discretetheory}

In this section we will consider systems of discrete equations
and develop a theory analogous to the continuous case results
reviewed in Paragraph~\ref{ODEtheory}.
It should be noted that discrete equations might not possess continuous limits.
Such discrete equations have no relation to discretizations of ODEs.
As in the previous section
we will assume summation over repeated indices.

Let us consider discrete equations with the dependent variable
$$
 {\bf u} _m = ( u^1 _m , ..., u^q _m  ) ,
 \qquad
 m \in \mathbb{Z}  .
$$
Discrete systems of order $n$ can be presented as equations involving $n+1$ points
\begin{equation} \label{difference1}
F_  {\beta}  ( m, {\bf u}_m , {\bf u}_{m+1}  ,{\bf u}_{m+2}  , ... , {\bf u}_{m+n}   ) = 0  ,
\qquad
 {\beta}  = 1, 2, ..., r .
\end{equation}
We will assume that these equations can be resolved for $ {\bf u}_m$  and $ {\bf u}_{m+n} $.
This assumption is necessary to solve the  Cauchy problem to the left and to the right from
the lattice points containing initial values.

We consider Lie point symmetries
\begin{equation}  \label{symmetry1a}
X = \eta ^{\alpha}  ( {\bf u} )  {\partial \over \partial  u  ^{\alpha}  } .
\end{equation}
For application to functions on lattice points we consider symmetry operators
which are prolonged to all points involved in equations~(\ref{difference1})
\begin{equation}  \label{symmetry1b}
X = \eta ^{\alpha} _m {\partial \over \partial  u  ^{\alpha} _m }
 + \eta ^{\alpha} _{m+1} {\partial \over \partial  u  ^{\alpha} _{m+1}  }
 + ...
 +  \eta ^{\alpha} _{m+n} {\partial \over \partial  u  ^{\alpha} _{m+n}  }   ,
\qquad
\eta ^{\alpha} _l = \eta ^{\alpha} ( {\bf u} _l ) .
\end{equation}
It is helpful  to introduce forward and backwards shift operators $S_+$ and $S_-$:
$$
S_+ m = m+1, \quad  S_+ {\bf u}_ m =  {\bf u} _{m+1} ,
$$
$$
S_- m = m-1, \quad  S_- {\bf u}_ m =  {\bf u} _{m-1} .
$$

Discrete variational operators are defined by the relation
\begin{multline*}
\delta \sum _m {\cal F}  ( m,  {\bf u}_m ,  {\bf u}_{m+1}  ,  ... , {\bf  u}_{m+n}   )
\\
= \sum _m    \delta {u} ^{\alpha} _m
\sum _{k = 0} ^{\infty} S_- ^k  {\partial \over \partial  u ^{\alpha} _{m+k} }
 {\cal F}  ( m,  {\bf u}_m ,   {\bf u}_{m+1}  ,  ... , {\bf  u}_{m+n} ) .
\end{multline*}
We suppose ${\cal F} \rightarrow 0 $ sufficiently fast  when $ m
\rightarrow \pm \infty$ so that the difference functional is well
defined.
The relation provides us with operators
\begin{equation}  \label{variational1}
{ \delta   \over \delta u_m ^{\alpha}}
= \sum _{k = 0} ^{\infty} S_- ^k   {\partial \over \partial  u ^{\alpha} _{m+k} }
=   { \partial \over \partial u_m ^{\alpha}  }
+  S_-   { \partial \over \partial u _{m+1} ^{\alpha} }
+ ...
+  S_- ^k  { \partial  \over \partial u _{m+k} ^{\alpha} }
+ ...
\end{equation}
Note that these operators are given for the system~(\ref{difference1})
of arbitrary order $n$.

We will make use of adjoint variables
$ {\bf v}_m = ( v^1 _m, ..., v^r  _m ) $
and adjoint equations
\begin{equation}  \label{adjoint1}
F _{\alpha} ^* = { \delta   \over \delta u_m ^{\alpha}   } ( v_m ^ {\beta}   F _ {\beta}   ) = 0 ,
\qquad {\alpha} = 1, ..., q ,
\end{equation}
which are always linear for the adjoint variables $ {\bf v}_m$.
These equations can be presented as
\begin{multline*}
F _{\alpha} ^*
= v_m ^{\beta} { \partial F _{\beta} \over \partial u_m ^{\alpha}  }
+ v_{m-1} ^{\beta} S_-  \left( { \partial F _{\beta} \over \partial u _{m+1} ^{\alpha} } \right)
+ ...
\\
+ v_{m-k} ^{\beta} S_- ^k  \left( { \partial F _{\beta} \over \partial u _{m+k} ^{\alpha} } \right)
+ ...
+ v_{m-n} ^{\beta} S_- ^n  \left( { \partial F _{\beta} \over \partial u _{m+n} ^{\alpha}  } \right) = 0 .
\end{multline*}

\begin{remark}
The system which consists of the original equations~(\ref{difference1})
together with the adjoint equations~(\ref{adjoint1})
can be considered as variational equations for the formal Lagrangian
\begin{equation}   \label{discreteformal1}
L =  v_m ^{\beta} F  _ {\beta}   .
\end{equation}
\end{remark}

Now we will obtain the main identity which will be used to find first integrals.

\medskip

Let us fix the value of index $m$,
which corresponds to the left point in the equations~(\ref{difference1}),
and define {\it higher} order  discrete Euler--Lagrange operators
\begin{equation} \label{higher1}
{ \delta    \over \delta  u  _{m (j)} ^{\alpha}  }
= \sum _{k = 0 } ^{\infty}  S_- ^{k}  { \partial   \over \partial u  _{m+j+k} ^ {\alpha}   }
= { \partial   \over \partial u  _{m+j} ^ {\alpha}  }
+ S_- { \partial   \over \partial u  _{m+j+1} ^ {\alpha}  }
+ ...
+ S_- ^k  { \partial   \over \partial u  _{m+j+k} ^ {\alpha} }
+ ...
\end{equation}
We note that variational operators~(\ref{variational1}) belong to this family:
\begin{equation*}
{ \delta    \over \delta  u  _{m}  ^ {\alpha} }
= { \delta    \over \delta  u  _{m (0)} ^ {\alpha}  } .
\end{equation*}

\begin{lemma}  \label{ideintity1} {\bf (Main identity for discrete equations)}
The following identity holds
\begin{equation}     \label{ideintity1a}
v   _m   ^ {\beta}   X   F _ {\beta}
=  \eta  _m ^ {\alpha}    F^* _ {\alpha}    +   ( 1 - S_- ) J  ,
\end{equation}
where
\begin{equation}
J = \sum _{j = 1 } ^{n}
 \eta  _{m+j} ^{\alpha}    { \delta   \over \delta u_{m (j)} ^ {\alpha}    }
 ( v  _m ^ {\beta}  F _ {\beta}    )
\end{equation}
\end{lemma}

\noindent {\it Proof.}
The identity can be shown by a direct calculation.
\hfill $\Box$
\medskip

An alternative derivation of the main identity~(\ref{ideintity1a})
can be based on the following operator identity.

\begin{lemma}  \label{lemma11}
The following operator identity (no summation for ${\alpha}$)  holds
\begin{equation}   \label{discreteoperator}
\sum _{k= 0} ^{\infty}    \eta  _{m+k} ^ {\alpha}    { \partial    \over \partial  u_{m+k} ^ {\alpha}   }
=  \eta  ^ {\alpha}  _{m}
\sum _{k= 0} ^{\infty}   S_- ^k      { \partial    \over \partial  u_{m+k}  ^{\alpha}   }
 +  ( 1 - S_- )
 \sum _{j = 1 } ^{\infty}
 \eta  _{m+j} ^ {\alpha}    { \delta   \over \delta u_{m (j)} ^ {\alpha}    }
\end{equation}
\end{lemma}

If we take summation of the identities~(\ref{discreteoperator})
for all $ {\alpha}  = 1, ..., q $ and apply the resulting operator identity
to the formal Lagrangian~(\ref{discreteformal1}),
we get the identity~(\ref{ideintity1a}).

Let us adapt the results of the continuous case,
given in Paragraph~\ref{ODEtheory},
to the discrete case.

\begin{theorem}   \label{result1}  {\bf (Main theorem for discrete equations)}
Let the adjoint equations~(\ref{adjoint1})
be satisfied  for all solutions of the original equations~(\ref{difference1})
upon a substitution
\begin{equation}  \label{substitition1}
{ \bf v}_m = { \bf \varphi } ( m, { \bf u}_m ) ,
\qquad
{ \bf \varphi }    {\not\equiv} 0 .
\end{equation}
Then, any Lie point symmetry~(\ref{symmetry1a}) of the equations~(\ref{difference1})
leads to first integral
\begin{equation}  \label{first1}
J = \sum _{j = 1 } ^{n}
 \eta  _{m+j}  ^ {\alpha}   { \delta   \over \delta u_{m (j)} ^ {\alpha}    }
 ( v  _m ^ {\beta}  F _ {\beta}   )  ,
\end{equation}
where values ${\bf v}_m$, ..., ${\bf v}_{m-n}$  should be eliminated
be means of the equations~(\ref{substitition1}) and their shifts to the left.
\end{theorem}

\noindent {\it Proof.}
The result follows from the identity~(\ref{ideintity1a}).
\hfill $\Box$
\medskip

\begin{remark}   \label{reduction}
Generally first integrals $J$, given by ~(\ref{first1}), can depend on more than $n$ points.
We will call such expression {\it higher} first integrals.
Using the discrete equations~(\ref{difference1}),
we can always reduce this number of points to minimal set,
for example, to points $m, \ m+1, \ ...,  \ m+n-1$,
i.e.,  $\tilde{J} ( m, {\bf u}_m , {\bf u}_{m+1}  ,{\bf u}_{m+2}  , ... , {\bf u}_{m+n-1}   ) $.
\end{remark}

\begin{remark}
Instead of  point substitutions~(\ref{substitition1})
we can use generalized substitutions which involve neighbouring points.
For systems~(\ref{difference1}) we can consider substitutions like
\begin{equation}
{ \bf v}_m = { \bf \varphi } ( m, { \bf u}_m , { \bf u}_ {m+1} ) ,
\qquad   ... ,
\qquad
{ \bf v}_m = { \bf \varphi } ( m, { \bf u}_m , { \bf u}_ {m+1} , ... ,  { \bf u}_ {m+n-1}   ) .
\end{equation}
\end{remark}

\begin{remark}
The requirement that the substitution~(\ref{substitition1}) annihilates the adjoint equations
on the solutions of the original equations,
which is used in Theorem~\ref{result1},
can be replaced by a weaker condition
\begin{equation*}
\eta ^ {\alpha}  _m    F^* _ {\alpha}     = 0 ,
\end{equation*}
which should hold for a given symmetry $X$  of the system~(\ref{difference1})
on the solutions of this system.
This is a weaker condition than the requirement of the theorem
that all equations $ F^* _ {\alpha}  =0 $ hold individually.
\end{remark}

In the following sections we will consider applications of these results.

\section{Case of mapping involving a single dependent variable}

\label{discrete}

\subsection{General theory}

\label{single}

In this section we will consider scalar discrete equations
of order $n$
\begin{equation} \label{difference2}
F  ( m, u_m ,  u_{m+1}  ,    u_{m+2}  , ... ,   u_{m+n}   ) = 0
\end{equation}
admitting symmetries of the form
\begin{equation}  \label{symmetry2a}
X = \eta (u)  {\partial \over \partial  u } .
\end{equation}
Such symmetries are prolonged as
\begin{equation}  \label{symmetry2b}
X = \eta _{m} {\partial \over \partial  u _{m} }
+ \eta _{m+1} {\partial \over \partial  u _{m+1} }
+ ...
+ \eta _{m+n} {\partial \over \partial  u _{m+n} },
\qquad
\eta  _{l} = \eta ( u _{l} )
\end{equation}
to all points involved in the equation~(\ref{difference2}).

The corresponding adjoint equation~(\ref{adjoint1}) has the form
\begin{equation}  \label{adjoint2}
F ^* = { \delta   \over \delta u_m  } ( v_m F  ) = 0  ,
\end{equation}
where
\begin{equation}  \label{variationalscalar}
{ \delta   \over \delta u_m }
= \sum _{k = 0} ^{\infty} S_- ^k   {\partial \over \partial  u  _{m+k} }
=   { \partial \over \partial u_m  }
+  S_-   { \partial \over \partial u _{m+1}  }
+ ...
+  S_- ^k  { \partial  \over \partial u _{m+k}  }
+ ...
\end{equation}
is the discrete variational operator.
Explicitly we have
\begin{multline*}
F ^*
= v_m  { \partial F  \over \partial u_m   }
+ v_{m-1}  S_-  \left( { \partial F  \over \partial u _{m+1}  } \right)
+ ...
\\
+ v_{m-k}  S_- ^k  \left( { \partial F  \over \partial u _{m+k}  } \right)
+ ...
+ v_{m-n}  S_- ^n  \left( { \partial F  \over \partial u _{m+n}   } \right) = 0 .
\end{multline*}

Theorem~\ref{result1}  restricted to the case of this section states the following.

\begin{theorem}   \label{result2}
{\bf (Main theorem for scalar discrete equations)}
Let the adjoint equation~(\ref{adjoint2})
be satisfied  for all solutions of the original equation~(\ref{difference2})
upon a substitution
\begin{equation}  \label{substitition2}
v_m = \varphi ( m, u_m ) ,
\qquad
\varphi    {\not\equiv} 0 .
\end{equation}
Then, any Lie point symmetry~(\ref{symmetry2a}) of the equation~(\ref{difference2})
leads to a first integral
\begin{equation}
J = \sum _{j = 1 } ^{n}
 \eta  _{m+j}   { \delta   \over \delta u_{m (j)}   }
 ( v  _m F  )  ,
\end{equation}
where
$$
{ \delta    \over \delta  u  _{m (j)}  }
= \sum _{k = 0 } ^{\infty}  S_- ^{k}  { \partial   \over \partial u  _{m+j+k}    }
= { \partial   \over \partial u  _{m+j}   }
+ S_- { \partial   \over \partial u  _{m+j+1}   }
+ ...
+ S_- ^k  { \partial   \over \partial u  _{m+j+k}  }
+ ...
$$
and values $v_m$, ..., $v_{m-n}$ should be eliminated
by means of the Eq.~(\ref{substitition2}) and its shifts to the left.
\end{theorem}

\subsection{Four-point scalar discrete equation and example}

\label{four-point}

Let us specify the general results of the previous paragraph
to one scalar four-point discrete equation
\begin{equation*}
F  ( m , u_m ,  u_{m+1}  ,    u_{m+2} ,  u_{m+3}  ) = 0 .
\end{equation*}
The invariance condition reads as follows
\begin{equation*}
\left. X (F )  \right| _{F = 0 }
=  \left[
\eta _{m} {\partial F \over \partial  u _{m} }
+ \eta _{m+1} {\partial F \over \partial  u _{m+1} }
+ \eta _{m+2} {\partial F \over \partial  u _{m+2} }
+ \eta _{m+3} {\partial F \over \partial  u _{m+3} }
\right] _{F = 0 }
= 0 .
\end{equation*}

The adjoint equation takes the form
\begin{equation*}  \label{adjoint5}
F ^*
= v_m {\partial F \over \partial  u _{m} }
+  v_{m-1} S_- \left( {\partial F \over \partial  u _{m+1} }  \right)
+  v_{m-2} S_- ^2 \left( {\partial F \over \partial  u _{m+2} }  \right)
+  v_{m-3} S_- ^3 \left( {\partial F \over \partial  u _{m+3} }  \right)
= 0 .
\end{equation*}
The first integral gets presented as
\begin{multline*}
J = \eta _{m+1}
\left[     v_m  { \partial F \over \partial  u _{m+1} }
 +   S_- \left( v_m {\partial F \over \partial  u _{m+2} }  \right)
 +   S_- ^2 \left( v_m {\partial F \over \partial  u _{m+3} }  \right)  \right]
\\
+  \eta _{m+2}
\left[     v_m  { \partial F \over \partial  u _{m+2} }
 + S_- \left( v_m {\partial F \over \partial  u _{m+3} }  \right)  \right]
+  \eta _{m+3}
\left[     v_m  { \partial F \over \partial  u _{m+3} }  \right] .
\end{multline*}

\bigskip

\noindent {\bf Example}

\medskip

Let us consider the discrete equation
\begin{equation}   \label{discrete2e}
F  = {  ( u_{m+3} -  u_{m+1} ) (  u_{m+2} -  u_{m} )   \over
             ( u_{m+3} -  u_{m+2} ) (  u_{m+1} -  u_{m} )  }  - K      = 0   , \qquad   K \neq 0 .
\end{equation}
This equation was considered in~\cite{[Pavel], [Pavel2]}
as a part of the system~(\ref{discrete3e}),
which will be examined below.

We note that for $K = 0$  this equation is equivalent to the system
$$
\begin{array}{l}
 u_{m+2} -  u_{m} = 0 , \\
 \\
 u_{m+1} -  u_{m} \neq 0 ,\\
\end{array}
$$
which can be easily solved as
$$
u_m = A (-1)^m  + B ,
\qquad
A \neq 0 .
$$

The equation~(\ref{discrete2e}) admits symmetries
\begin{equation}   \label{symmetri3}
X_1 = {\frac{ \partial }{\partial u}} ,
\qquad
X_2 = u {\frac{ \partial }{\partial u}} ,
\qquad
X_3 = u^2 {\frac{ \partial }{\partial u}} .
\end{equation}

The adjoint equation~(\ref{adjoint2})
(after use of the original equation $F= 0$)
is
\begin{equation}     \label{adjointm}
F^*  = { K (  u_{m+2} -  u_{m+1}) \over (  u_{m+2} -  u_{m}) (  u_{m+1} -  u_{m+1}) }
( v_{m}  + (1-K)  v_{m-1}   + (K-1) v_{m-2}  -  v_{m-3} )  = 0 .
\end{equation}
It gets simplified to a linear mapping
\begin{equation}    \label{linear}
v_{m}  + (1-K)  v_{m-1}   + (K-1) v_{m-2}  -  v_{m-3}   = 0 .
\end{equation}

It is easy to find solutions $  v_{m} = v_{m} (m) $.
We obtain three independent solutions of the adjoint equation
\begin{equation}
\begin{array}{llll}
K  = 4 :
 &
v_m ^a = 1 , &    v _m ^b  = m , & v _m ^c = m^2   ; \\
0<K  \ \ \mbox{or} \ \  K>4 :
 &
v _ m ^a  = 1  , &    v _ m ^b  = \mu _{1}   ^m    , & v _ m ^c  = \mu _{2}   ^m    ;     \\
0<K<4 :
 &
v _m ^a  = 1 , &    v_m ^b  = \cos ( 2 \phi m  ) , & v _m ^c =  \sin ( 2 \phi m  )  ;  \\
\end{array}
\end{equation}
where
\begin{equation}
\mu _{1,2} =  { (K-2) \pm \sqrt{K^2 - 4K } \over 2 }
\qquad
\mbox{and}
\qquad
\phi = \arccos \left( {\sqrt{K}  \over 2}  \right) .
\end{equation}

First of all we consider the solution of the adjoint equation  $ v_ m ^a = 1 $,
which is common for all values $K \neq 0 $.
Applying Theorem~\ref{result1} with this solution and symmetries $X_1$, $X_2$ and $X_3$
and simplifying the obtained first integrals as described in Remark~\ref{reduction},
we get the first integrals
\begin{multline*}
\tilde{J}_{1a} = 2  \left(
{ K  \over u_{m+2} - u_{m} }
- { 1 \over u_{m+2} - u_{m+1} }
- { 1 \over u_{m+1} - u_{m} }
\right) ,
\\
\tilde{J}_{2a}
= { K ( u_{m+2} + u_{m} )  \over u_{m+2} - u_{m} }
-  { 2 u_{m+1} \over u_{m+2} - u_{m+1} }
 -   { 2 u_{m+1} \over u_{m+1} - u_{m} } ,
\\
\tilde{J}_{3a} =
2 \left(
 { K u_{m+2} u_{m}  \over u_{m+2} - u_{m} }
-  { u_{m+1} ^2  \over u_{m+2} - u_{m+1} }
 -    { u_{m+1} ^2 \over u_{m+1} - u_{m} }
 \right) ,
\end{multline*}
respectively.

These three first integrals, which hold for all $K \neq 0$, are not independent.
They satisfy the relation
\begin{equation}
\tilde{J}_{1a}  \tilde{J}_{3a}  - ( \tilde{J}_{2a} ) ^2 = 4 K - K^2 .
\end{equation}
To integrate the discrete equation~(\ref{discrete2e})
we need one more independent first integral
(it should be a first integral which involves $m$).
As in the continuous case
we need to consider different cases of the parameter $K$ separately.

\bigskip

\noindent {\bf Case:  $K = 4$}

\medskip

For $K = 4$ the other solutions of the adjoint equation  are
$$
v_ m ^b = m
\qquad
\mbox{and}
\qquad
v_ m ^c = m^2  .
$$

For $v_ m ^b=m $  and symmetries~(\ref{symmetri3})
we get first integrals
\begin{multline*}
\tilde{J}_{1b} = 2 m \left[
{ K  \over u_{m+2} - u_{m} }
- { 1 \over u_{m+2} - u_{m+1} }
- { 1 \over u_{m+1} - u_{m} }
\right]
\\
-  K  \left(
  { 1  \over u_{m+2} - u_{m} }
+ { 1\over u_{m+1} - u_{m} }
\right)
+ { 3 \over u_{m+2} - u_{m+1} }
+ { 3 \over u_{m+1} - u_{m} } ,
\end{multline*}
\begin{multline*}
\tilde{J}_{2b} =  m \left[
\frac{  K( u_{m+2} + u_m ) }{u_{m+2} - u_m}
-  { 2  u_{m+1}  \over  u_{m+2} - u_{m+1} }
-  { 2  u_{m+1}  \over  u_{m+1} - u_{m} }
\right]
\\
-  { K \over 2 } \left(
  { u_{m+2} + u_{m} \over u_{m+2} - u_{m} }
+ { u_{m+1} + u_{m} \over u_{m+1} - u_{m} }
\right)
+ { 3 u_{m+1} \over u_{m+2} - u_{m+1} }
+ { 3 u_{m+1} \over u_{m+1} - u_{m} }   ,
\end{multline*}
\begin{multline*}
\tilde{J}_{3b} = 2  m  \left[
\frac{ K u_{m+2} u_m}{u_{m+2} - u_m}
-  {   u_{m+1} ^2 \over  u_{m+2} - u_{m+1} }
-  {   u_{m+1} ^2 \over  u_{m+1} - u_{m} }
\right]
\\
-   K  \left(
  { u_{m+2}  u_{m} \over u_{m+2} - u_{m} }
+ { u_{m+1}  u_{m} \over u_{m+1} - u_{m} }
\right)
+ { 3 u_{m+1} ^2 \over u_{m+2} - u_{m+1} }
+ { 3 u_{m+1} ^2 \over u_{m+1} - u_{m} }   .
\end{multline*}

For $ v_ m ^c =m^2 $ we get first integrals
\begin{multline*}
\tilde{J}_{1c} = 2  m^2    \left[
{ K  \over u_{m+2} - u_{m} }
- { 1 \over u_{m+2} - u_{m+1} }
- { 1 \over u_{m+1} - u_{m} }
\right] \\
 +  m \left[
-  2 K  \left(
  { 1  \over u_{m+2} - u_{m} }
+ { 1  \over u_{m+1} - u_{m} }
\right)
+ { 6 \over u_{m+2} - u_{m+1} }
+ { 6 \over u_{m+1} - u_{m} }
\right] \\
+  K \left(
  { 1  \over u_{m+2} - u_{m} }
+ { 1\over u_{m+1} - u_{m} }
\right)
- { 5 \over u_{m+2} - u_{m+1} }
- { 5 \over u_{m+1} - u_{m} }  ,
\end{multline*}
\begin{multline*}
\tilde{J}_{2c} = m^2 \left[
\frac{ K (u_m + u_{m+2})}{u_{m+2} - u_m}
- { 2 u_{m+1}  \over u_{m+2} - u_{m+1} }
- { 2 u_{m+1}  \over u_{m+1} - u_{m} }
\right]
\\
  + m \left[
-  K  \left(
  { u_{m+2} + u_{m} \over u_{m+2} - u_{m} }
+ { u_{m+1} + u_{m} \over u_{m+1} - u_{m} }
\right)
+ { 6 u_{m+1} \over u_{m+2} - u_{m+1} }
+ { 6 u_{m+1} \over u_{m+1} - u_{m} }
  \right]
 \\
+  {K \over 2}  \left(
  { u_{m+2} +  u_{m}  \over u_{m+2} - u_{m} }
+ { u_{m+1} +  u_{m}  \over u_{m+1} - u_{m} }
\right)
- { 5 u_{m+1} \over u_{m+2} - u_{m+1} }
- { 5 u_{m+1} \over u_{m+1} - u_{m} }   ,
\end{multline*}
\begin{multline*}
\tilde{J}_{3c} = 2 m^2  \left[
\frac{ K u_{m+2} u_m }{u_{m+2} - u_m}
- {  u_{m+1} ^2 \over u_{m+2} - u_{m+1} }
- {  u_{m+1} ^2 \over u_{m+1} - u_{m} }
\right]
\\
+ m \left[
- 2   K   \left(
  { u_{m+2}  u_{m}  \over u_{m+2} - u_{m} }
+ { u_{m+1}  u_{m}  \over u_{m+1} - u_{m} }
\right)
+ { 6 u_{m+1} ^2 \over u_{m+2} - u_{m+1} }
+ { 6 u_{m+1} ^2 \over u_{m+1} - u_{m} }
\right]
 \\
+  K   \left(
  { u_{m+2}  u_{m}  \over u_{m+2} - u_{m} }
+ { u_{m+1}  u_{m}  \over u_{m+1} - u_{m} }
\right)
- { 5 u_{m+1} ^2 \over u_{m+2} - u_{m+1} }
- { 5 u_{m+1} ^2 \over u_{m+1} - u_{m} }  .
\end{multline*}

In total we obtained nine nontrivial first integrals.
They satisfy 6 relations
$$
\tilde{J}_{1a} \tilde{J}_{3a} - \tilde{J}_{2a}  ^2 = 0   ,
\qquad
\tilde{J}_{1b} \tilde{J}_{3b} - \tilde{J}_{2b}  ^2 = - 4   ,
\qquad
\tilde{J}_{1c} \tilde{J}_{3c} - \tilde{J}_{2c}  ^2 =   4  ,
$$
$$
\tilde{J}_{1a} \tilde{J}_{1c} - \tilde{J}_{1b}  ^2 -  {1 \over 4} \tilde{J}_{1a}  ^2 =  0   ,
\qquad
\tilde{J}_{2a} \tilde{J}_{2c} - \tilde{J}_{2b}  ^2 -  {1 \over 4} \tilde{J}_{2a}  ^2 = - 4   ,
\qquad
\tilde{J}_{3a} \tilde{J}_{3c} - \tilde{J}_{3b}  ^2 -  {1 \over 4} \tilde{J}_{3a}  ^2= 0   .
$$
As we know,  for a four-point equation we have at most three independent first integrals

\bigskip

\noindent {\bf Integration of the mapping}

\medskip

Let us chose three first integrals
$\tilde{J}_{1a}$, $\tilde{J}_{2a}$ and $\tilde{J}_{1b}$.
The Jacobian is
$$
J = \mbox{det} \left(
{ \partial  (  \tilde{J}_{1a}, \tilde{J}_{2a}, \tilde{J}_{1b} )  \over
 \partial  ( u_m, u_{m+1} , u_{m+2} )  } \right)
=  { 16 ( u_{m+2} - 2 u_{m+1} + u_{m} ) ^4
\over
( u_{m+1} - u_{m} ) ^3    ( u_{m+2} - u_{m} ) ^3 ( u_{m+2} - u_{m+1}  ) ^3   }
$$

\begin{enumerate}

\item

For $J \neq 0 $ we can set these first integrals equal to constants,
eliminate $u_{m+1}$ and $u_{m+2}$ from them
and express $u_m $ in terms of $m$ and the constants.
We obtain
\begin{equation}
u_m = { 1 \over C_1 m + C_2 } + C_3 ,
\end{equation}
where $C_1 \neq 0$,  $C_2$ and  $C_3$ are constants.

\item

For $J =  0 $ we solve the system
$$
\begin{array}{l}
u_{m+2} - 2 u_{m+1} + u_{m} = 0 , \\
\\
u_{m+1} \neq  u_{m} , \qquad u_{m+2} \neq  u_{m} \\
\end{array}
$$
and obtain its solution
\begin{equation}  \label{sub1a}
u_m  = C_1 m + C_2 , \qquad C_1 \neq 0  .
\end{equation}
Using substitution into  the discrete equation~(\ref{discrete2e}),
we confirm that~(\ref{sub1a}) is a solution.
It is a degenerate solution since it depends only on two constants.

\end{enumerate}

\bigskip

\noindent {\bf Case:  $ K < 0 $ or $K > 4$}

\medskip

In this case
two specific solutions of the adjoint equation~(\ref{adjointm})  are
$$
v_ m ^b = \mu _{1} ^m
\qquad
\mbox{and}
\qquad
v_ m ^c = \mu _{2} ^m   ,
\qquad
\mu _{1,2} = { (K-2) \pm \sqrt{K^2 - 4K } \over 2 } .
$$


For $ v_ m ^{b,c} =   \mu _i  ^m $
($i = 1$ for $b$ and $i = 2$ for $c$)
we obtain first integrals

\begin{multline*}
\tilde{J}_{1b,1c}
= \mu_i^{m} \left[
   { K  \over u_{m+2} - u_{m} }
-  { K  \over u_{m+1} - u_{m} }
\right]
\\
+ \mu_i^{m-1}  \left[
 K \left(
   { 1 \over u_{m+2} - u_{m} }
+  { 1  \over u_{m+1} - u_{m} }
\right)
-  { 1 \over u_{m+2} - u_{m+1} }
-  { 1  \over u_{m+1} - u_{m} }
\right]  \\
- \mu_i^{m-2}  \left[
   { 1  \over u_{m+2} - u_{m+1} }
+  { 1  \over u_{m+1} - u_{m} }
\right]   ,
\end{multline*}
\begin{multline*}
\tilde{J}_{2b,2c}
= \mu_i^{m} \left[
   { K  u_{m}  \over u_{m+2} - u_{m} }
-  { K  u_{m}  \over u_{m+1} - u_{m} }
\right]
\\
+ \mu_i^{m-1}  \left[
 {K \over 2}  \left(
   { u_{m+2} + u_{m} \over u_{m+2} - u_{m} }
+  { u_{m+1} + u_{m}  \over u_{m+1} - u_{m} }
\right)
-  {  u_{m+1} \over u_{m+2} - u_{m+1} }
-  {  u_{m+1} \over u_{m+1} - u_{m} }
\right]  \\
- \mu_i^{m-2}  \left[
   {  u_{m+1}  \over u_{m+2} - u_{m+1} }
+  {  u_{m+1}  \over u_{m+1} - u_{m} }
\right]   ,
\end{multline*}
\begin{multline*}
\tilde{J}_{3b,3c}
= \mu_i^{m} \left[
   { K  u_{m} ^2  \over u_{m+2} - u_{m} }
-  { K  u_{m} ^2  \over u_{m+1} - u_{m} }
\right]
\\
+ \mu_i^{m-1}  \left[
 K   \left(
   { u_{m+2}  u_{m} \over u_{m+2} - u_{m} }
+  { u_{m+1}  u_{m}  \over u_{m+1} - u_{m} }
\right)
-  {  u_{m+1} ^2 \over u_{m+2} - u_{m+1} }
-  {  u_{m+1} ^2 \over u_{m+1} - u_{m} }
\right]  \\
- \mu_i^{m-2}  \left[
   {  u_{m+1} ^2  \over u_{m+2} - u_{m+1} }
+  {  u_{m+1} ^2  \over u_{m+1} - u_{m} }
\right]    .
\end{multline*}
These first integrals correspond to three symmetries~(\ref{symmetri3}),  respectively.

The first integrals of this case
obey the relations
$$
\tilde{J}_{1a} \tilde{J}_{3a} - \tilde{J}_{2a}  ^2 = K (4-K)  ,
\qquad
\tilde{J}_{1b} \tilde{J}_{3b} - \tilde{J}_{2b}  ^2 = 0  ,
\qquad
\tilde{J}_{1c} \tilde{J}_{3c} - \tilde{J}_{2c}  ^2 = 0  ,
$$
$$
\tilde{J}_{1b}   \tilde{J}_{1c}   - {K \over 4}  \tilde{J}_{1a} ^2  = 0 ,
\qquad
\tilde{J}_{2b}   \tilde{J}_{2c}   - {K \over 4}  \tilde{J}_{2a} ^2  = {K^2 (4-K) \over 4}  ,
\qquad
\tilde{J}_{3b}   \tilde{J}_{3c}   - {K \over 4}  \tilde{J}_{3a} ^2  = 0
$$

\bigskip

\noindent {\bf Integration of the mapping}

\medskip

Let us chose
$\tilde{J}_{1a}$, $\tilde{J}_{2a}$ and $\tilde{J}_{1b} $
as three first integrals and find the Jacobian
$$
J = \mbox{det} \left(
{ \partial (  \tilde{J}_{1a}, \tilde{J}_{2a}, \tilde{J}_{1b} )  \over
 \partial  ( u_m, u_{m+1} , u_{m+2} ) } \right)
 =   K ( \mu_1 - 1 )  \mu_1 ^{m-2}
$$
$$
\times
{  K   ( u_{m+2} - 2 u_{m+1} + u_{m} )^2 +  (4-K)  ( u_{m+2} - u_{m} )^2
\over
( u_{m+1} - u_{m} ) ^3    ( u_{m+2} - u_{m} ) ^3 ( u_{m+2} - u_{m+1}  ) ^3   }
( \mu_1 u_{m+2} - ( \mu_1 + 1)  u_{m+1} + u_{m} ) ^2  .
$$

\begin{enumerate}

\item

For $J \neq 0$ we set first integrals equal to constants and obtain
\begin{equation}     \label{long}
u_m =
 C_1
{ (4-K) ( \mu _2 - \mu _1)
- C_2  ( 1 - \mu _1)^2  \mu _2 ^m
+ {K \over C_2 }   ( 1 - \mu _2)^2  \mu _1 ^m
\over
 K ( \mu _2 - \mu _1)
- C_2  ( 1 - \mu _1 ^2 )  \mu _2 ^m
+ {K \over C_2  }   ( 1 - \mu _2 ^2 ) \mu _1 ^m  }
+ C_3 ,
\end{equation}
where $C_1 \neq 0$,  $C_2 \neq 0 $ and  $C_3$ are constants.

\item

The case $J = 0$ splits into two subcases.

\begin{enumerate}

\item

The system
$$
\begin{array}{l}
K    ( u_{m+2} - 2 u_{m+1} + u_{m} )^2 +  (4-K)  ( u_{m+2} - u_{m} )^2 = 0, \\
\\
u_{m+1} \neq  u_{m},
\qquad
u_{m+2} \neq  u_{m}
\end{array}
$$
leads to
$$
\begin{array}{l}
{\displaystyle
u_{m+2} - 2 u_{m+1} + u_{m}  = \pm \sqrt{ K-4 \over K}  ( u_{m+2} - u_{m} )  } , \\
\\
u_{m+1} \neq  u_{m},
\qquad
u_{m+2} \neq  u_{m}
\end{array}
$$
and provides us with the degenerate solutions
\begin{equation}   \label{rere}
u_m  = C_1  \mu_1 ^m  + C_2
\qquad
\mbox{and}
\qquad
u_m  = C_1  \mu_2 ^m  + C_2 ,
\qquad
C_1 \neq 0 .
\end{equation}

\item

The system
$$
\begin{array}{l}
\mu_1 u_{m+2} - ( \mu_1 + 1)  u_{m+1} + u_{m}  = 0 , \\
\\
u_{m+1} \neq  u_{m},
\qquad
u_{m+2} \neq  u_{m}
\end{array}
$$
leads to
$$
u_m  = C_1  \mu_1 ^{-m}  + C_2 =  C_1  \mu_2 ^{m}  + C_2 ,
\qquad
C_1 \neq 0 ,
$$
which is the second of the two degenerate solutions~(\ref{rere}) already obtained.

\end{enumerate}

\end{enumerate}

The generic solution~(\ref{long}) can be conveniently rewritten as follows

\begin{itemize}

\item

$K > 4$:

\begin{equation}
u_m = C_1     \tanh ( \psi  m  + C_2 ) + C_3
\end{equation}
or
\begin{equation}
u_m = C_1   \coth ( \psi  m  + C_2 ) + C_3
\end{equation}

\item

$K < 0  $:

\begin{equation}
u_m =
\left\{
\begin{array}{l}
 C_1     \tanh ( \psi  m  + C_2 ) + C_3  \qquad \mbox{if $m$ is even} \\
 \\
 C_1     \coth ( \psi  m  + C_2 ) + C_3  \qquad \mbox{if $m$ is odd} \\
\end{array}
\right.
\end{equation}
or
\begin{equation}
u_m =
\left\{
\begin{array}{l}
 C_1     \coth ( \psi  m  + C_2 ) + C_3  \qquad \mbox{if $m$ is even} \\
 \\
 C_1     \tanh ( \psi  m  + C_2 ) + C_3  \qquad \mbox{if $m$ is odd} \\
\end{array}
\right.
\end{equation}

\end{itemize}

Here
$$
 \psi
 = { 1 \over 2 } \ln |  \mu_1 |
 = { 1 \over 2 } \ln \left| { K - 2  +  \sqrt{K^2 - 4K} \over 2 } \right|
$$
and $C_1 \neq 0$,  $C_2  $ and  $C_3$ are constants.

In addition to the generic solutions there are the degenerate solutions
\begin{equation}
u_m  = C_1  \mu_1 ^m  + C_2
\qquad
\mbox{and}
\qquad
u_m  = C_1  \mu_2 ^m  + C_2 ,
\qquad
C_1 \neq 0 ,
\end{equation}
which can be rewritten as
\begin{equation}
u_m  = C_1 ( \sgn K ) ^m   e^{\pm   2 \psi m } + C_2  .
\end{equation}

\bigskip

\noindent {\bf Case:  $  0 < K <  4$}

\medskip

In this case we obtain two specific solutions of the adjoint equation~(\ref{adjointm})
$$
v_ m ^b = \cos(2 \phi m )
\qquad
\mbox{and}
\qquad
v_ m ^c = \sin(2 \phi m )  ,
\qquad
\phi = \arccos \left( {\sqrt{K}  \over 2}  \right) .
$$

Application of Theorem~\ref{result1} with symmetries~(\ref{symmetri3})
and solution   $v_ m ^{b} $ gives us the first integrals
\begin{multline*}
\tilde{J}_{1b}
= \cos ( 2 \phi m )  \left[
   { K  \over u_{m+2} - u_{m} }
-  { K  \over u_{m+1} - u_{m} }
\right]
\\
+ \cos ( 2 \phi ( m - 1 ) )   \left[
 K \left(
   { 1 \over u_{m+2} - u_{m} }
+  { 1  \over u_{m+1} - u_{m} }
\right)
-  { 1 \over u_{m+2} - u_{m+1} }
-  { 1  \over u_{m+1} - u_{m} }
\right]  \\
- \cos ( 2 \phi ( m - 2 ) )  \left[
   { 1  \over u_{m+2} - u_{m+1} }
+  { 1  \over u_{m+1} - u_{m} }
\right]   ,
\end{multline*}
\begin{multline*}
\tilde{J}_{2b}
= \cos ( 2 \phi m ) \left[
   { K  u_{m}  \over u_{m+2} - u_{m} }
-  { K  u_{m}  \over u_{m+1} - u_{m} }
\right]
\\
+ \cos ( 2 \phi ( m - 1 ) )  \left[
 {K \over 2}  \left(
   { u_{m+2} + u_{m} \over u_{m+2} - u_{m} }
+  { u_{m+1} + u_{m}  \over u_{m+1} - u_{m} }
\right)
-  {  u_{m+1} \over u_{m+2} - u_{m+1} }
-  {  u_{m+1} \over u_{m+1} - u_{m} }
\right]  \\
- \cos ( 2 \phi ( m - 2 ) ) \left[
   {  u_{m+1}  \over u_{m+2} - u_{m+1} }
+  {  u_{m+1}  \over u_{m+1} - u_{m} }
\right]   ,
\end{multline*}
\begin{multline*}
\tilde{J}_{3b}
= \cos ( 2 \phi m )  \left[
   { K  u_{m} ^2  \over u_{m+2} - u_{m} }
-  { K  u_{m} ^2  \over u_{m+1} - u_{m} }
\right]
\\
+ \cos ( 2 \phi ( m - 1 ) )  \left[
 K   \left(
   { u_{m+2}  u_{m} \over u_{m+2} - u_{m} }
+  { u_{m+1}  u_{m}  \over u_{m+1} - u_{m} }
\right)
-  {  u_{m+1} ^2 \over u_{m+2} - u_{m+1} }
-  {  u_{m+1} ^2 \over u_{m+1} - u_{m} }
\right]  \\
- \cos ( 2 \phi ( m - 2 ) )  \left[
   {  u_{m+1} ^2  \over u_{m+2} - u_{m+1} }
+  {  u_{m+1} ^2  \over u_{m+1} - u_{m} }
\right]    .
\end{multline*}

For $v_ m ^{c}$ we get the first integrals
\begin{multline*}
\tilde{J}_{1c}
= \sin ( 2 \phi m )  \left[
   { K  \over u_{m+2} - u_{m} }
-  { K  \over u_{m+1} - u_{m} }
\right]
\\
+ \sin ( 2 \phi ( m - 1 ) )   \left[
 K \left(
   { 1 \over u_{m+2} - u_{m} }
+  { 1  \over u_{m+1} - u_{m} }
\right)
-  { 1 \over u_{m+2} - u_{m+1} }
-  { 1  \over u_{m+1} - u_{m} }
\right]  \\
- \sin ( 2 \phi ( m - 2 ) )  \left[
   { 1  \over u_{m+2} - u_{m+1} }
+  { 1  \over u_{m+1} - u_{m} }
\right]   ,
\end{multline*}
\begin{multline*}
\tilde{J}_{2c}
= \sin ( 2 \phi m ) \left[
   { K  u_{m}  \over u_{m+2} - u_{m} }
-  { K  u_{m}  \over u_{m+1} - u_{m} }
\right]
\\
+ \sin ( 2 \phi ( m - 1 ) )  \left[
 {K \over 2}  \left(
   { u_{m+2} + u_{m} \over u_{m+2} - u_{m} }
+  { u_{m+1} + u_{m}  \over u_{m+1} - u_{m} }
\right)
-  {  u_{m+1} \over u_{m+2} - u_{m+1} }
-  {  u_{m+1} \over u_{m+1} - u_{m} }
\right]  \\
- \sin ( 2 \phi ( m - 2 ) ) \left[
   {  u_{m+1}  \over u_{m+2} - u_{m+1} }
+  {  u_{m+1}  \over u_{m+1} - u_{m} }
\right]   ,
\end{multline*}
\begin{multline*}
\tilde{J}_{3c}
= \sin ( 2 \phi m )  \left[
   { K  u_{m} ^2  \over u_{m+2} - u_{m} }
-  { K  u_{m} ^2  \over u_{m+1} - u_{m} }
\right]
\\
+ \sin ( 2 \phi ( m - 1 ) )  \left[
 K   \left(
   { u_{m+2}  u_{m} \over u_{m+2} - u_{m} }
+  { u_{m+1}  u_{m}  \over u_{m+1} - u_{m} }
\right)
-  {  u_{m+1} ^2 \over u_{m+2} - u_{m+1} }
-  {  u_{m+1} ^2 \over u_{m+1} - u_{m} }
\right]  \\
- \sin ( 2 \phi ( m - 2 ) )  \left[
   {  u_{m+1} ^2  \over u_{m+2} - u_{m+1} }
+  {  u_{m+1} ^2  \over u_{m+1} - u_{m} }
\right]    .
\end{multline*}

The first integrals of this case
together with first integrals
$\tilde{J}_{1a} $, $\tilde{J}_{2a}$ and $\tilde{J}_{3a} $
satisfy the relations
$$
\tilde{J}_{1a} \tilde{J}_{3a} - \tilde{J}_{2a}  ^2 = K ( 4-K)  ,
\qquad
\tilde{J}_{1b} \tilde{J}_{3b} - \tilde{J}_{2b}  ^2 = {K^2 (K-4) \over 4} ,
\qquad
\tilde{J}_{1c} \tilde{J}_{3c} - \tilde{J}_{2c}  ^2 = {K^2 (K-4) \over 4}  ,
$$
$$
\tilde{J}_{1b}  ^2 + \tilde{J}_{1c} ^2  - {K \over 4}  \tilde{J}_{1a} ^2 = 0 ,
\qquad
\tilde{J}_{2b}  ^2 + \tilde{J}_{2c} ^2  - {K \over 4}  \tilde{J}_{2a} ^2 = {K^2 (4-K) \over 4}   ,
\qquad
\tilde{J}_{3b}  ^2 + \tilde{J}_{3c} ^2  - {K \over 4}  \tilde{J}_{3a} ^2 = 0 .
$$

\bigskip

\noindent {\bf Integration of the mapping}

\medskip

As in the previous case we chose three first integrals
$ \tilde{J}_{1a}$, $ \tilde{J}_{2a}$ and $ \tilde{J}_{1b}  $.
The Jacobian is
$$
J = \mbox{det} \left(
{ \partial (  \tilde{J}_{1a}, \tilde{J}_{2a}, \tilde{J}_{1b}  )  \over
 \partial  ( u_m, u_{m+1} , u_{m+2} )  } \right)
$$
$$
 =
{  K    ( u_{m+2} - 2 u_{m+1} + u_{m} )^2 +  (4-K)  ( u_{m+2} - u_{m} )^2
\over
( u_{m+1} - u_{m} ) ^3    ( u_{m+2} - u_{m} ) ^3 ( u_{m+2} - u_{m+1}  ) ^3   }
{ K R_1 R_2 \over \cos( 2 \phi ( m+1) ) - \cos(  2 \phi m   )  } ,
$$
where
$$
R_1 =  \alpha     ( u_{m+2} - 2 u_{m+1} + u_{m} ) +  \beta  ( u_{m+2} - u_{m} )  ,
$$
$$
\alpha = \sin 2 \phi \left(  \sin (2 \phi m )  + \sin  \phi  \right) ,
\qquad
\beta  = ( 1 -  \cos 2 \phi ) \left(  \cos (2 \phi m )  - \cos  \phi  \right)
$$
and
$$
R_2 =  \gamma    ( u_{m+2} - 2 u_{m+1} + u_{m} ) +  \delta ( u_{m+2} - u_{m} )  ,
$$
$$
\gamma = \sin 2 \phi \left(  \sin (2 \phi m )  - \sin  \phi  \right) ,
\qquad
\delta   = ( 1 -  \cos 2 \phi ) \left(  \cos (2 \phi m )  + \cos  \phi  \right)  .
$$

\begin{enumerate}

\item

In the case $J \neq 0$ we set these first integrals equal to constants
and obtain the generic solution
\begin{equation}
u_m =   C_1  \tan ( \phi  m  + C_2  ) + C_3  ,
\end{equation}
where $C_1 \neq 0$,
$C_2  \neq - {3 \over 2} \phi + { \pi \over 2} k $, $k   \in \mathbb{Z}  $
and  $C_3$ are constants.

\item

Analysis of the  case $J = 0$ splits into three subcases.

\begin{enumerate}

\item

The system
$$
\begin{array}{l}
 K    ( u_{m+2} - 2 u_{m+1} + u_{m} )^2 +  (4-K)  ( u_{m+2} - u_{m} )^2 = 0 , \\
\\
  u_{m+1} \neq  u_{m} ,
\qquad
 u_{m+2} \neq  u_{m} \\
\end{array}
$$
has no solutions.

\item

The system
$$
\begin{array}{l}
R_1 = 0 , \\
\\
  u_{m+1} \neq  u_{m} ,
\qquad
 u_{m+2} \neq  u_{m} \\
\end{array}
$$
has solutions
\begin{equation}
u_m =   C_1  \tan ( \phi  m  + C_2  ) + C_3  ,
\qquad
C_1 \neq 0 ,
\quad
C_2  =  - {3 \over 2} \phi + { \pi k } .
\end{equation}
Verification shows that these functions are solutions
of the discrete equations~(\ref{discrete2e}).

\item

The system
$$
\begin{array}{l}
R_2 = 0 , \\
\\
  u_{m+1} \neq  u_{m} ,
\qquad
 u_{m+2} \neq  u_{m} \\
\end{array}
$$
has solutions
\begin{equation}
u_m =   C_1  \tan ( \phi  m  + C_2  ) + C_3  ,
\qquad
C_1 \neq 0 ,
\quad
C_2  =  - {3 \over 2} \phi +   {\pi  \over 2}   +  { \pi k } ,
\end{equation}
which are solutions of the discrete equation.

\end{enumerate}

\end{enumerate}

Finally, we unite the obtained solutions into the generic solution of the form
\begin{equation}
u_m =   C_1  \tan ( \phi  m  + C_2  ) + C_3  ,
\end{equation}
where $C_1 \neq 0$, $C_2 $ and  $C_3$ are  constants.

\section{Integrating factors for mappings}

\label{integrating2}

In this section we consider the direct method for discrete equations.
It is an adaptation of the continuous case method.
We would like to find first integrals
\begin{equation}
I (m, u_m, u_{m+1}, ...,u_{m+n-1}) =  \mbox{const}
\end{equation}
which hold on the solutions of the scalar discrete equation
\begin{equation}
F(m, u_m, u_{m+1}, ...,u_{m+n}) = 0 .
\end{equation}
we require
\begin{equation} \label{direct_delta_Psi}
(S_+ - 1 )  I = \Lambda  F  ,
\qquad
\Lambda  = \Lambda (m, u_m, u_{m+1}, ...,u_{m+n}) .
\end{equation}

\begin{remark}
In the continuous case it was useful to assume that the equation is linear
with respect to the highest derivative (see Remark~\ref{linearity}).
This lead to integrating factors independent of the  highest derivative.
In the discrete case there is no analog of this property and
a discrete integrating factor can depend on the same variables as the discrete equation.
This makes application of the direct method for discrete equations more difficult.
One needs to come up with a good Ansatz for the integration factor $ \Lambda $
in order to determine it.
\end{remark}

Relation~(\ref{direct_delta_Psi}) should hold identically,
i.e., not only on the solutions of the discrete equation.
Since the left hand side is a total difference
we can apply the discrete variational operator~(\ref{variationalscalar}) to it
and obtain the equation
\begin{equation} \label{direct_delta_2}
\frac{\delta}{\delta u_m } ( \Lambda   F ) = 0
\end{equation}
for $ \Lambda $.

Finally we note that
all solutions of equation~(\ref{direct_delta_2})
always satisfy the  discrete adjoint equation
\begin{equation}
\left.
\frac{\delta}{\delta u_m } ( v  _m  F )
\right|_{F = 0 }
= 0 :
\end{equation}
For integrating factors equation~(\ref{direct_delta_2}) holds identically,
while solutions of the adjoint equation hold on the solutions of the original discrete equation.

\bigskip

\noindent {\bf Example}

\medskip

We demonstrate the direct method on the example of the discrete equation
\begin{equation} \label{int_factors_delta_eqn}
F(u_m, u_{m+1}, u_{m+2}, u_{m+3})
= \frac{(u_{m+3}-u_{m+1})(u_{m+2}-u_{m})}{(u_{m+3}-u_{m+2})(u_{m+1}-u_{m})} - K = 0,
\qquad
K  = \mbox{const}
\end{equation}

In general, the integrating factors for~(\ref{int_factors_delta_eqn}) have the following form
\begin{equation}
\Lambda(m, u_m, u_{m+1}, u_{m+2}, u_{m+3}).
\end{equation}
Equation~(\ref{direct_delta_2}) becomes
$$
\left(
{ \partial \over \partial u_m   }
+  S_-   { \partial \over \partial u _{m+1}  }
+  S_- ^2  { \partial \over \partial u _{m+2}  }
+  S_- ^3  { \partial \over \partial u _{m+3}  }
\right)  ( \Lambda   F ) = 0
$$
or in detail
\begin{multline} \label{direct_delta_F_detail}
{ ( u_{m+3} - u_{m+1}) ( u_{m+2} - u_{m} )  \over ( u_{m+3} - u_{m+2} ) ( u_{m+1} - u_{m} ) }
\, \frac{\partial \Lambda}{\partial u_m}
+ {   ( u_{m+2} - u_{m} ) ( u_{m+1} - u_ {m-1} ) \over ( u_{m+2} - u_ {m+1} ) ( u_{m} - u_ {m-1}  ) }
\, \frac{\partial \Lambda^-}{\partial u_m}
\\
+ { ( u_{m+1} - u_{m-1} ) ( u_{m} - u_{m-2} ) \over ( u_{m+1} -u_{m} ) ( u_{m-1} - u_{m-2} ) }
\, \frac{\partial \Lambda^{--}}{\partial u_m}
+ { ( u_{m} - u_{m-2} ) ( u_{m-1} - u_{m-3} ) \over ( u_{m} -u_{m-1} ) ( u_{m-2} - u_{m-3} ) }
\, \frac{\partial \Lambda^{---}}{\partial u_m}
\\
+ { ( u_{m+3} - u_{m+1}) ( u_{m+2} - u_{m+1} )  \over ( u_{m+3} - u_{m+2} ) ( u_{m+1} - u_{m} ) ^2 }
\, \Lambda
- {   ( u_{m+2} - u_{m-1} ) ( u_{m+1} - u_ {m-1} ) \over ( u_{m+2} - u_ {m+1} ) ( u_{m} - u_ {m-1}  ) ^2 }
\, \Lambda^-
\\
+ { ( u_{m+1} - u_{m-1} ) ( u_{m+1} - u_{m-2} ) \over ( u_{m+1} -u_{m} ) ^2 ( u_{m-1} - u_{m-2} ) }
\, \Lambda^{--}
- { ( u_{m-1} - u_{m-2} ) ( u_{m-1} - u_{m-3} ) \over ( u_{m} -u_{m-1} )^2  ( u_{m-2} - u_{m-3} ) }
\, \Lambda^{---}
\\
 - K \frac{\partial }{\partial u_m} \,
\left(
\Lambda + \Lambda^- + \Lambda^{--} + \Lambda^{---}
\right) = 0 ,
\end{multline}
where
$ \Lambda^- = S_-  \Lambda   $, $ \Lambda^{--} = S_- ^2  \Lambda  $ and $ \Lambda^{---} = S_- ^3 \Lambda $.

Even to find a special solution of the last equation is a very difficult task.
Practically, we have to introduce some kind of Ansatz to find a solution.
If some integrating factor is obtained, we can use the equation
\begin{equation}
(S_+ - 1 )  I (m, u_m, u_{m+1},u_{m+2})   = \Lambda  F
\end{equation}
to find the corresponding first integral.
We will not pursue this issue here.
Instead we will show how the direct method can provide the same first integrals
which were given in the previous section, where equation~(\ref{int_factors_delta_eqn}) was solved.

Let us try
\begin{equation} \label{lambda_ansatz_1}
\Lambda = f_1(m, u_m,u_{m+1},u_{m+2}) + f_2(m, u_{m+1},u_{m+2},u_{m+3}).
\end{equation}
as an Ansatz.
Among such functions there are integration factors
\begin{multline} \label{lambdas_family}
\Lambda
= \left(
{ 1 \over  u_{m+2} - u_{m+1} }   - { 1 \over  u_{m+2} - u_{m} }
\right)
(a_2 (m) u_{m+2}^2 + a_1 (m) u_{m+2} + a_0 (m) )
\\
+  \left(
{ 1 \over  u_{m+3} - u_{m+1} }   - { 1 \over  u_{m+2} - u_{m+1} }
\right)
(a_2 (m+1) u_{m+1}^2 + a_1 (m+1) u_{m+1} + a_0 (m+1) ) ,
\end{multline}
where $a_2(m)$, $a_1(m)$ and $a_0(m) $ are solutions of the equation
\begin{equation} \label{adj_eqn}
a_i (m) + (1-K)  a_i (m-1) + (K-1) a_i (m-2) - a_i (m-3) = 0,
\quad i = 0, 1, 2.
\end{equation}
(Substitution of $\Lambda$ in the form~(\ref{lambdas_family})
into the equation~(\ref{direct_delta_F_detail}) gives equations~(\ref{adj_eqn})
for functions $a_2(m)$, $a_1(m)$ and $a_0(m) $.)
Note that this is the same equation as equation~(\ref{linear}),
which was studied before.
So, we have the same solution.

Using integrating factor Ansatz~(\ref{lambdas_family}),
we can obtain the same first integrals as given in Paragraph~\ref{four-point}.
Specification of the functions $ a_0(m)$, $  a_1(m) $ and $  a_2(m)$
which provides these first integrals is given in Table~1.
Here we use three linearly independent solutions of the equation~(\ref{adj_eqn}):
a constant solution and functions
\begin{equation} \label{alpha}
\alpha (m)
=
\left\{
\begin{array}{ll}
-( m + (m - 1) ),  & \mbox{if} \   K = 4 ,\\
-( \mu _1 ^m + \mu _1 ^{m - 1}  )  & \mbox{if} \   K < 0 \ \mbox{or} \  K > 4 , \\
-( \cos( 2 \phi m )  + \cos( 2 \phi (m - 1)  ) ) &   \mbox{if} \   0 < K < 4 , \\
\end{array}
\right.
\end{equation}
and
\begin{equation} \label{beta}
\beta (m)
=
\left\{
\begin{array}{ll}
-( m^2 + (m - 1)^2 ) & \mbox{if} \    K = 4 , \\
-( \mu _2 ^m + \mu _2 ^{m - 1}  ) & \mbox{if} \    K < 0 \ \mbox{or} \  K > 4 ,  \\
-( \sin( 2 \phi m )  + \sin( 2 \phi (m - 1)  ) ) & \mbox{if} \     0 < K < 4 ,   \\
\end{array}
\right.
\end{equation}
where
$$
\mu _{1,2} =  { (K-2) \pm \sqrt{K^2 - 4K } \over 2 }
\qquad
\mbox{and}
\qquad
 \phi = \arccos \left( {\sqrt{K}  \over 2}  \right)  .
$$

\begin{figure}
{

\begin{center}
{ \bf Table 1.} First integrals corresponding to integrating factors~(\ref{lambdas_family})
with specified functions $ a_0(m)$, $  a_1(m) $ and $  a_2(m)$.
Functions $\alpha (m)$  and $\beta (m)$ are given in~(\ref{alpha}) and~(\ref{beta}), respectively.
\end{center}

$$
\begin{array}{|c|c|c|c|}
\hline
 &   &   &   \\
  a_0(m) &  a_1(m)  &   a_2(m)  & \mbox{first integral} \\
 &   &   &   \\
\hline
  &   &   &   \\
-2 &  0 &  0 &   \tilde{J} _{1 a} \\
 &   &   &   \\
0 & -2 & 0 &  \tilde{J} _{2 a} \\
 &   &   &   \\
0 & 0 & -2 & \tilde{J} _{3 a} \\
 &   &   &   \\
\alpha (m)  &  0 &  0 &   \tilde{J} _{1 b} \\
 &   &   &   \\
0 & \alpha (m)  & 0 &  \tilde{J} _{2 b} \\
 &   &   &   \\
0 & 0 & \alpha (m)  & \tilde{J} _{3 b} \\
 &   &   &   \\
 \beta (m) &  0 &  0 &   \tilde{J} _{1 c} \\
 &   &   &   \\
0 & \beta (m) & 0 &  \tilde{J} _{2 c} \\
 &   &   &   \\
0 & 0 & \beta (m)  & \tilde{J} _{3 c} \\
 &   &   &   \\
\hline
\end{array}
$$

}
\end{figure}

\section{Discretizations of a scalar ODE}

\label{discretization}

\subsection{Theory for difference systems}

\label{fordifference}

In this section we are interested in dicretizations of a scalar
ODE. For the discretization of an ODE of order $n$ we
need a difference stencil with at least $n+1$ points. We will use
precisely $n+1$ points, namely, points $    x _m $, ...,  $ x_{m+n}$.
These points are not specified in advance and will be defined by
an additional mesh equation~\cite{[D-book]}.

As a discretization we will consider a discrete equation on $n+1$ points
\begin{equation} \label{difference3}
F  ( x _m ,  u_m , x_{m+1} ,  u_{m+1}  ,  ... , x_{m+n} ,  u_{m+n}   ) = 0 ,
\end{equation}
on a mesh
\begin{equation}  \label{mesh}
 \Omega   ( x _m ,  u_m , x_{m+1} ,  u_{m+1}   , ... , x_{m+n} ,  u_{m+n}   ) = 0 .
\end{equation}
These two equations form the difference system to be used. In the
continuous limit the first equation goes into the original ODE and
the second equation turns into  an identity (for example, $0=0$).

The Lie point symmetry
\begin{equation}  \label{symmetry3}
X
= \xi  (x,u) {\partial \over \partial  x  }
+ \eta (x,u) {\partial \over \partial  u  }
\end{equation}
gets prolonged to the points of the difference stencil as
\begin{equation}
X
= \xi  _{m} {\partial \over \partial  x _{m} }
+ \eta _{m} {\partial \over \partial  u _{m} }
+ ...
+ \xi  _{m+n} {\partial \over \partial  x _{m+n} }
+ \eta _{m+n} {\partial \over \partial  u _{m+n} } ,
\end{equation}
$$
 \xi  _{l} = \xi ( x _{l} , u _{l} ) ,
\qquad
 \eta  _{l} = \eta ( x _{l} , u _{l} )  .
$$

The discrete variational operators~(\ref{variational1}) take the form
\begin{equation}   \label{variational3a}
{ \delta   \over \delta u_m  }
= \sum _{k = 0} ^{\infty} S_- ^k   {\partial \over \partial  u _{m+k} }
=   { \partial \over \partial u_m  }
+  S_-   { \partial \over \partial u _{m+1}}
+ ...
+  S_- ^k  { \partial  \over \partial u _{m+k}}
+ ... ,
\end{equation}
\begin{equation}   \label{variational3b}
{ \delta   \over \delta x_m  }
= \sum _{k = 0} ^{\infty} S_- ^k   {\partial \over \partial  x _{m+k} }
=   { \partial \over \partial x_m  }
+  S_-   { \partial \over \partial x _{m+1}}
+ ...
+  S_- ^k   { \partial  \over \partial x _{m+k}}
+ ...
\end{equation}
To the system of difference equations~(\ref{difference3}),(\ref{mesh})
there correspond the adjoint equations
\begin{equation}  \label{adjoint3a}
F ^* = { \delta   \over \delta u_m  } ( v_m F + w_m \Omega ) = 0
\end{equation}
and
\begin{equation} \label{adjoint3b}
\Omega ^* = { \delta   \over \delta x_m  } ( v_m F + w_m \Omega ) =  0  ,
\end{equation}
where  $ v_m $  and $ w_m $ are adjoint variables.
In detail they are
\begin{multline*}
F ^* = v_m  { \partial F \over \partial u_m  }
+ v_{m-1} S_-  \left( { \partial F \over \partial u _{m+1}} \right)
+ ...
+ v_{m-k} S_- ^k  \left( { \partial F \over \partial u _{m+k}} \right)
+ ...
+ v_{m-n} S_- ^n  \left( { \partial F \over \partial u _{m+n}} \right)
\\
+ w_m  { \partial \Omega \over \partial u_m  }
+ w_{m-1} S_-  \left( { \partial \Omega \over \partial u _{m+1}} \right)
+ ...
+ w_{m-k} S_- ^k  \left( { \partial \Omega \over \partial u _{m+k}} \right)
+ ...
+ w_{m-n} S_- ^n  \left( { \partial \Omega \over \partial u _{m+n}} \right)
=  0
\end{multline*}
and
\begin{multline*}
\Omega ^*  = v_m  { \partial F \over \partial x_m  }
+ v_{m-1} S_-  \left( { \partial F \over \partial x _{m+1}} \right)
+ ...
+ v_{m-k} S_- ^k  \left( { \partial F \over \partial x _{m+k}} \right)
+ ...
+ v_{m-n} S_- ^n  \left( { \partial F \over \partial x _{m+n}} \right)
\\
+ w_m  { \partial \Omega \over \partial x_m  }
+ w_{m-1} S_-  \left( { \partial \Omega \over \partial x _{m+1}} \right)
+ ...
+ w_{m-k} S_- ^k  \left( { \partial \Omega \over \partial x _{m+k}} \right)
+ ...
+ w_{m-n} S_- ^n  \left( { \partial \Omega \over \partial x _{m+n}} \right)
= 0 .
\end{multline*}

In this setting Theorem~\ref{result1} takes the following form

\begin{theorem}   \label{result3}
{\bf (Main result for discretized ODE)}
Let the adjoint equations~(\ref{adjoint3a}),(\ref{adjoint3b})
be  satisfied  for all solutions of the original equations~(\ref{difference3}),(\ref{mesh})
upon a substitution
\begin{equation}  \label{substitition3}
\begin{array}{l}
v_m = \varphi _1 ( m,x_m,u_m ) , \\
\\
w_m = \varphi _2 ( m,x_m,u_m ) , \\
\end{array}
\qquad
\qquad
\varphi _1    {\not\equiv} 0
\quad
\mbox{or}
\quad
\varphi _2  {\not\equiv} 0 .
\end{equation}
Then, any Lie point symmetry~(\ref{symmetry3}) of the equations~(\ref{difference3}),(\ref{mesh})
leads to first integral
\begin{equation}    \label{f_int_ODE}
J = \sum _{j = 1 } ^{n}
\left(
\xi  _{m+j}   { \delta   \over \delta x_{m (j)}   }
+ \eta  _{m+j}   { \delta   \over \delta u_{m (j)}   }
\right)
 ( v  _m F   +  w_m  \Omega )
 ,
\end{equation}
where
\begin{equation} \label{higher3a}
{ \delta    \over \delta  u  _{m (j)}  }
= \sum _{k = 0 } ^{\infty}  S_- ^{k}  { \partial   \over \partial u  _{m+j+k}  }
= { \partial   \over \partial u  _{m+j}  }
+ S_- { \partial   \over \partial u  _{m+j+1}  }
+ ...
+ S_- ^k  { \partial   \over \partial u  _{m+j+k}  }
+ ...
\end{equation}
and
\begin{equation} \label{higher3b}
{ \delta    \over \delta  x  _{m (j)}   }
= \sum _{k = 0 } ^{\infty}  S_- ^{k}  { \partial   \over \partial x  _{m+j+k}  }
= { \partial   \over \partial x  _{m+j}  }
+ S_- { \partial   \over \partial x  _{m+j+1}  }
+ ...
+ S_- ^k  { \partial   \over \partial x  _{m+j+k}  }
+ ...
\end{equation}
are higher order discrete Euler--Lagrange operators
and $v_m$, $w_m$, ...,   $v_{m-n}$, $w_{m-n}$  should be eliminated
by means of Eqs.~(\ref{substitition3}) and their shifts to the left.
\end{theorem}

\begin{remark}   \label{helpforODE}
As in the general case the condition that the adjoint equations are satisfied,
i.e.,  $ F^* = \Omega ^* = 0 $,
can be substituted by a weaker condition
\begin{equation*}
\xi _m  \Omega ^* + \eta _m  F^* = 0 ,
\end{equation*}
which should hold for a given symmetry $X$  of the system~(\ref{difference3}),(\ref{mesh})
on the solutions of this system.
\end{remark}


\subsection{Discretization of second order ODEs}

\label{forsecondorder}

Let us specify the formulas of the previous section
for the three-point case.
We get the difference equation
\begin{equation*}
F  ( x _m ,  u_m , x_{m+1} ,  u_{m+1}  ,  x_{m+2} ,  u_{m+2}    ) = 0
\end{equation*}
on the mesh
\begin{equation*}
 \Omega   ( x _m ,  u_m , x_{m+1} ,  u_{m+1}  ,  x_{m+2} ,  u_{m+2}    ) = 0 .
\end{equation*}

The adjoint equations take the form
\begin{multline*}
F ^* = v_m  { \partial F \over \partial u_m  }
+ v_{m-1} S_-  \left( { \partial F \over \partial u _{m+1}} \right)
+ v_{m-2} S_- ^2  \left( { \partial F \over \partial u _{m+2}} \right)
\\
+ w_m  { \partial \Omega \over \partial u_m  }
+ w_{m-1} S_-  \left( { \partial \Omega \over \partial u _{m+1}} \right)
+ w_{m-2} S_- ^2  \left( { \partial \Omega \over \partial u _{m+2}} \right)
=  0
\end{multline*}
and
\begin{multline*}
\Omega ^*  = v_m  { \partial F \over \partial x_m  }
+ v_{m-1} S_-  \left( { \partial F \over \partial x _{m+1}} \right)
+ v_{m-2} S_- ^2  \left( { \partial F \over \partial x _{m+2}} \right)
\\
+ w_m  { \partial \Omega \over \partial x_m  }
+ w_{m-1} S_-  \left( { \partial \Omega \over \partial x _{m+1}} \right)
+ w_{m-2} S_- ^2  \left( { \partial \Omega \over \partial x _{m+2}} \right)
= 0 .
\end{multline*}

First integrals are given as
\begin{multline*}
J =
\left[
\xi _{m+1}
\left(      { \partial  \over \partial  x _{m+1} }
 +   S_-   {\partial   \over \partial  x _{m+2}  }  \right)
+ \eta _{m+1}
\left(      { \partial  \over \partial  u _{m+1} }
 +   S_-   {\partial   \over \partial  u _{m+2}  }  \right)
 \right.
\\
 \left.
+  \xi _{m+2}
 { \partial  \over \partial  x _{m+2} }
+  \eta _{m+2}
 { \partial  \over \partial  u _{m+2} }
 \right]
  ( v  _m F   +  w_m  \Omega ) .
\end{multline*}

\bigskip

\noindent {\bf Example: Harmonic oscillator}

\medskip

Let us consider the one-dimensional harmonic oscillator
\begin{equation} \label{oscil}
 \ddot{u}  + u = 0 .
\end{equation}
As a discretization we consider the scheme
\begin{equation}
 { 2 \over x_{m+2} -  x_m   }
 \left(
   { u_{m+2} - u_{m+1}  \over x_{m+2} - x_{m+1}  }
 - { u_{m+1} - u_{m}  \over x_{m+1} - x_{m}  }
 \right)
 +  { u_{m+2}  + 2 u_{m+1}  + u_m  \over 4 }  = 0
\end{equation}
on the uniform mesh
\begin{equation}
x_{m+2} - x_{m+1} = x_{m+1} - x_{m}  .
\end{equation}
This discretization of the harmonic oscillator was considered in~\cite{[DK-3]}.

Let us rewrite the scheme in an equivalent form
\begin{equation}  \label{Lagrrangian}
\begin{array}{c}
{\displaystyle
F =  { u_{m+2} - u_{m+1}  \over x_{m+2} - x_{m+1}  }
 - { u_{m+1} - u_{m}  \over x_{m+1} - x_{m}  }
 +  { x_{m+2} - x_{m}  \over  2  } { u_{m+2}  + 2 u_{m+1}  + u_m  \over 4 }  = 0
 , } \\
\\
{\displaystyle
\Omega =  ( x_{m+2} - x_{m+1} ) - (  x_{m+1} - x_{m} )  = 0    .   } \\
\end{array}
\end{equation}
It is not difficult to verify that the difference system~(\ref{Lagrrangian})
admits the symmetries generated by the operators
\begin{equation}  \label{sym3}
X_1 = {\frac{ \partial }{\partial x}} ,
\qquad
X_2 = \sin(\omega x) {\frac{ \partial }{\partial u}} ,
\qquad
X_3 = \cos(\omega x) {\frac{ \partial }{\partial u}} ,
\qquad
X_4 = u {\frac{ \partial }{\partial u}} ,
\end{equation}
where
\begin{equation*}
\omega = {\frac{ \arctan ( h /2 ) }{h /2 }},
\qquad
h =  x_{m+2} - x_{m+1}  =    x_{m+1} - x_{m} .
\end{equation*}

The adjoint equations are
\begin{multline}  \label{ad3a}
F^* = v_m
\left(   { 1  \over x_{m+1} - x_{m}  }  +   { x_{m+2} - x_{m}  \over  8  }  \right)
\\
 +  v_{m-1}
\left(  -  { 1  \over x_{m+1} - x_{m}  }   -  { 1  \over x_{m} - x_{m-1}  } +   { x_{m+1} - x_{m-1}  \over  4  }  \right)
\\
 + v_{m-2}
\left(   { 1  \over x_{m} - x_{m-1}  }  +   { x_{m} - x_{m-2}  \over  8  }  \right) = 0
\end{multline}
and
\begin{multline}  \label{ad3b}
\Omega^*
= v_m
\left(  -  { u_{m+1} - u_{m}  \over ( x_{m+1} - x_{m} ) ^2   }   -  { u_{m+2}  + 2 u_{m+1}  + u_m  \over 8 }   \right)
\\
 + v_{m-1}
\left(    { u_{m+1} - u_{m}  \over ( x_{m+1} - x_{m} ) ^2   }    +  { u_{m} - u_{m-1}  \over ( x_{m} - x_{m-1} ) ^2   }   \right)
\\
 + v_{m-2}
\left(  -  { u_{m} - u_{m-1}  \over ( x_{m} - x_{m-1} ) ^2   }   +  { u_{m}  + 2 u_{m-1}  + u_{m-2}  \over 8 }   \right)
\\
 + w_m - 2 w_{m-1}  + w_{m-2} = 0
\end{multline}
considered on the solutions of the equations~(\ref{Lagrrangian}).

It is easy to check that
on the solutions of the equations~(\ref{Lagrrangian})
the adjoint equations~(\ref{ad3a}),(\ref{ad3b})  have the particular solution
\begin{equation}   \label{pair01}
v_m ^a = 0 ,
\qquad
w_m ^a = x_m .
\end{equation}

For symmetries~(\ref{symmetry3})  with $ \xi = 0 $ we can
consider the equation~(\ref{ad3a}) instead of the system~(\ref{ad3a}),(\ref{ad3b})
(see Remark~\ref{helpforODE}). In this case we find the special solution
\begin{equation}   \label{pair02}
v_m ^b = u_m  ,
\qquad
w_m ^b = 0 .
\end{equation}

Let us use these solutions to find first integrals with the help of
Theorem~\ref{result3} and symmetries~(\ref{sym3}).
We will bypass the higher first integrals and provide only the final results
for both pairs~(\ref{pair01}) and~(\ref{pair02}).

\begin{itemize}

\item  $v_m ^a = 0 $, $ w_m ^a= x_m $

Application of the theorem with symmetry $X_1$ gives first integral
\begin{equation}   \label{frolattice}
\tilde{J}_1 ^a = x_m  - x_{m+1}  =  - h   .
\end{equation}
The other symmetries provide trivial first integrals.

\item  $v_m ^b = u_m  $, $ w_m ^b = 0  $


For symmetries $X_2$,  $X_3$ and $X_4$ we obtain first integrals
\begin{equation}
\tilde{J}_2 ^b = \left( { 1 \over h} + { h \over 4} \right)
\left( - u_{m+1} \sin ( \omega x_{m+1} ) +     u_{m} \sin ( \omega x_{m+2}  ) \right)  ,
\end{equation}
\begin{equation}
\tilde{J}_3 ^b= \left( { 1 \over h} + { h \over 4} \right)
\left( - u_{m+1} \cos ( \omega x_{m+1} ) +     u_{m} \cos ( \omega x_{m+2}  ) \right)  ,
\end{equation}
\begin{equation}
\tilde{J}_4 ^b
= - h \left[
\left( { u_{m+1} - u_{m} \over h} \right)^2
+ \left( { u_{m+1} + u_{m} \over 2} \right)^2
\right]  ,
\end{equation}
where we used $ h = x_{m+1} -x_m $ and $ x_{m+2} = x_{m+1} + h  $.

\end{itemize}

Using values of the first integrals  $\tilde{J}_1 ^a$ $ \tilde{J}_2 ^b$ and $\tilde{J}_3 ^b$,
we can express the solution of the difference system in the form
\begin{equation}  \label{oscL1}
u_m  = A \cos ( \omega x_m  ) + B  \sin ( \omega x _m ) .
\end{equation}
The mesh for this solution
\begin{equation}  \label{oscL2}
x_m = x_0 + m h, \qquad m = 0, \pm 1, \pm 2, ...
\end{equation}
can be obtained by integration of the linear equation~(\ref{frolattice}).
Here $A$, $B$, $h>0$ and $x_0$ are constants.
Note that   $x_0$  appears from the integration of the linear equation~(\ref{frolattice}).

\subsection{Discretization of third order ODEs}

\label{forthirdorder}

Here we specify the general formulas
for the four-point case.
We get the difference equation
\begin{equation*}
F  ( x _m ,  u_m , x_{m+1} ,  u_{m+1}  ,  x_{m+2} ,  u_{m+2} , x_{m+3} ,  u_{m+3}    ) = 0
\end{equation*}
on the mesh
\begin{equation*}
 \Omega   ( x _m ,  u_m , x_{m+1} ,  u_{m+1}  ,  x_{m+2} ,  u_{m+2}    ,  x_{m+3} ,  u_{m+3} ) = 0 .
\end{equation*}

The adjoint equations are
\begin{multline*}
F ^* = v_m  { \partial F \over \partial u_m  }
+ v_{m-1} S_-  \left( { \partial F \over \partial u _{m+1}} \right)
+ v_{m-2} S_- ^2  \left( { \partial F \over \partial u _{m+2}} \right)
+ v_{m-3} S_- ^3  \left( { \partial F \over \partial u _{m+3}} \right)
\\
+ w_m  { \partial \Omega \over \partial u_m  }
+ w_{m-1} S_-  \left( { \partial \Omega \over \partial u _{m+1}} \right)
+ w_{m-2} S_- ^2  \left( { \partial \Omega \over \partial u _{m+2}} \right)
+ w_{m-3} S_- ^3  \left( { \partial \Omega \over \partial u _{m+3}} \right)
=  0
\end{multline*}
and
\begin{multline*}
\Omega ^*  = v_m  { \partial F \over \partial x_m  }
+ v_{m-1} S_-  \left( { \partial F \over \partial x _{m+1}} \right)
+ v_{m-2} S_- ^2  \left( { \partial F \over \partial x _{m+2}} \right)
+ v_{m-3} S_- ^3  \left( { \partial F \over \partial x _{m+3}} \right)
\\
+ w_m  { \partial \Omega \over \partial x_m  }
+ w_{m-1} S_-  \left( { \partial \Omega \over \partial x _{m+1}} \right)
+ w_{m-2} S_- ^2  \left( { \partial \Omega \over \partial x _{m+2}} \right)
+ w_{m-3} S_- ^3  \left( { \partial \Omega \over \partial x _{m+3}} \right)
= 0 .
\end{multline*}

First integrals take the form
\begin{multline*}
J =
\left[
\xi _{m+1}
\left(      { \partial  \over \partial  x _{m+1} }
 +   S_-   {\partial   \over \partial  x _{m+2}  }
 +   S_-  ^2  {\partial   \over \partial  x _{m+3}  } \right)
 \right.
\\
+ \eta _{m+1}
\left(      { \partial  \over \partial  u _{m+1} }
 +   S_-   {\partial   \over \partial  u _{m+2}  }
 +   S_-  ^2  {\partial   \over \partial  u _{m+3}  }   \right)
\\
 +  \xi _{m+2}
\left(  { \partial  \over \partial  x _{m+2} }
 +   S_-  {\partial   \over \partial  x _{m+3}  }  \right)
+  \eta _{m+2}
\left(  { \partial  \over \partial  u _{m+2} }
 + S_-    {\partial   \over \partial  u _{m+3}  }  \right)
\\
 \left.
  +  \xi _{m+3}
   {\partial   \over \partial  x _{m+3}  }
  +  \eta _{m+2}
   {\partial   \over \partial  u _{m+3}  }
\right]
  ( v  _m F   +  w_m  \Omega ) .
\end{multline*}

\bigskip

\noindent {\bf Example}

\medskip

Let us return to the ODE
\begin{equation}   \label{odePW}
 F =  { 1 \over \dot{u} ^2 } \left( \dot{u}  \dddot{u} - { 3 \over 2 }  \ddot{u} ^2  \right) - M  = 0 ,
\end{equation}
which we examined in the Paragraph~\ref{mainexample}.
We recall that in the general case it admits symmetries
\begin{equation}   \label{part1}
X_1 = {\frac{ \partial }{\partial u}} ,
\qquad
X_2 = u {\frac{ \partial }{\partial u}} ,
\qquad
X_3 = u^2 {\frac{ \partial }{\partial u}} ,
\qquad
X_4 = {\frac{ \partial }{\partial x}} .
\end{equation}
For $M = 0 $ there are additional symmetries
\begin{equation}  \label{part2}
X_5 = x {\frac{ \partial }{\partial x}} ,
\qquad
X_6 = x^2 {\frac{ \partial }{\partial x}} .
\end{equation}
We will consider these two cases separately.

\bigskip

\noindent {\bf Case: $M = 0 $}

\medskip

As a discretization we consider the invariant scheme
\begin{equation}   \label{discrete3e}
\begin{array}{c}
{\displaystyle
F =  { u_{m+3} -  u_{m+1}  \over x_{m+3} -  x_{m+1}  }
{ u_{m+2} -  u_{m}  \over x_{m+2} -  x_{m}  }
- { u_{m+3} -  u_{m+2}  \over x_{m+3} -  x_{m+2}  }
{ u_{m+1} -  u_{m}  \over x_{m+1} -  x_{m}  }   = 0 , } \\
\\
{\displaystyle
\Omega =  {  ( x_{m+3} -  x_{m+1} ) (  x_{m+2} -  x_{m} )   \over
             ( x_{m+3} -  x_{m+2} ) (  x_{m+1} -  x_{m} )  }  - K     = 0   ,  \quad K \neq 0  , } \\
\end{array}
\end{equation}
which was introduced in~\cite{[Pavel], [Pavel2]}.
It admits all six symmetries~(\ref{part1}),(\ref{part2}).

The adjoint system for the presented scheme is
$$
F^*
= { \alpha  (  u_{m+2} -  u_{m+1}) \over (  u_{m+2} -  u_{m}) (  u_{m+1} -  u_{m}) }
( v_{m}  + (1-K)  v_{m-1}   + (K-1) v_{m-2}  -  v_{m-3} )  = 0
$$
and
\begin{multline*}
\Omega ^*
=  - { \alpha  (  x_{m+2} -  x_{m+1}) \over (  x_{m+2} -  x_{m}) (  x_{m+1} -  x_{m}) }
( v_{m}  + (1-K)  v_{m-1}   + (K-1) v_{m-2}  -  v_{m-3} ) \\
+  { K (  x_{m+2} -  x_{m+1}) \over (  x_{m+2} -  x_{m}) (  x_{m+1} -  x_{m}) }
( w_{m}  + (1-K)  w_{m-1}   + (K-1) w_{m-2}  -  w_{m-3} )  = 0 ,
\end{multline*}
where
$$
\alpha
= { u_{m+3} -  u_{m+1}  \over x_{m+3} -  x_{m+1}  }
{ u_{m+2} -  u_{m}  \over x_{m+2} -  x_{m}  }
=  { u_{m+3} -  u_{m+2}  \over x_{m+3} -  x_{m+2}  }
{ u_{m+1} -  u_{m}  \over x_{m+1} -  x_{m}  }  .
$$
Variables $u_{m+3}$ and $x_{m+3}$ in the coefficient $\alpha$
should be expressed in term of the other variables involved in the scheme.

The adjoint equations lead to the system of linear mappings
$$
 v_{m}  + (1-K)  v_{m-1}   + (K-1) v_{m-2}  -  v_{m-3} = 0 ,
$$
$$
 w_{m}  + (1-K)  w_{m-1}   + (K-1) w_{m-2}  -  w_{m-3} = 0 .
$$
One can use pairs $( v_m , w_m )$ which solve this system to find first integrals.

However, it is more convenient to rewrite the scheme~(\ref{discrete3e})
in the  equivalent form
\begin{equation}     \label{discrete2b}
\begin{array}{c}
{\displaystyle
\tilde{F} = {  ( u_{m+3} -  u_{m+1} ) (  u_{m+2} -  u_{m} )   \over
             ( u_{m+3} -  u_{m+2} ) (  u_{m+1} -  u_{m} )  }  - K     = 0       , } \\
\\
{\displaystyle
\Omega =  {  ( x_{m+3} -  x_{m+1} ) (  x_{m+2} -  x_{m} )   \over
             ( x_{m+3} -  x_{m+2} ) (  x_{m+1} -  x_{m} )  }  - K     = 0   .   } \\
\end{array}
\end{equation}
Note that the system is symmetric under the interchange of $u$ and $x$.
We can use the results obtained for discrete equation~(\ref{discrete2e})
to integrate this scheme.
We need to consider different subcases for different values of $K$.

\begin{enumerate}

\item

$K = 4$

We obtain the solution
\begin{equation}   \label{sol3a}
u_m =   {  1 \over C_1  m  + C_2   } +  C_3
\qquad
\mbox{or}
\qquad
u_m =  C_1  m  + C_2
\end{equation}
on the mesh
\begin{equation}   \label{sol3b}
x_m =   {  1 \over C_4  m  + C_5   } +  C_6
\qquad
\mbox{or}
\qquad
x_m =  C_4  m  + C_5   ,
\end{equation}
where  $C_1 \neq 0$,   $C_2$, $C_3$,  $C_4 \neq 0$,   $C_5$ and $C_6$ are constants.

\item

$K>4$

We obtain the solution
\begin{equation}  \label{sol3c}
u_m =   C_1   \tanh  ( \psi  m  + C_2  ) + C_3
\end{equation}
or
\begin{equation}  \label{sol3cnew}
u_m =   C_1   \coth  ( \psi  m  + C_2  ) + C_3
\end{equation}
or
\begin{equation}  \label{sol3d}
u_m  = C_1  \mu_{1,2} ^m  + C_2
    =  C_1  e^{\pm  2 \psi m } + C_2
\end{equation}
on the mesh
\begin{equation}   \label{sol3e}
x_m = C_4     \tanh ( \psi  m  + C_5  ) + C_6
\end{equation}
or
\begin{equation}   \label{sol3enew}
x_m = C_4     \coth ( \psi  m  + C_5  ) + C_6
\end{equation}
or
\begin{equation}   \label{sol3f}
x_m  = C_4  \mu_{1,2} ^m  + C_5
     =  C_4 e^{\pm  2 \psi m } + C_5  ,
\end{equation}
where $C_1 \neq 0$,  $C_2 $,   $C_3$, $C_4 \neq 0$,  $C_5  $ and $C_6$
are constants.
Here
$$
\mu_{1,2} =  { K - 2   \pm   \sqrt{K^2 - 4K} \over 2 }
$$
and
$$
\psi
=  {1\over 2} \ln \mu_{1}
=   {1\over 2} \ln \left( { K - 2   +  \sqrt{K^2 - 4K} \over 2 } \right)  .
$$

\item

$K<0$

We obtain the solution
\begin{equation}  \label{sol3cc}
u_m =
\left\{
\begin{array}{l}
 C_1     \tanh ( \psi  m  + C_2 ) + C_3  \qquad \mbox{if $m$ is even} \\
 \\
 C_1     \coth ( \psi  m  + C_2 ) + C_3  \qquad \mbox{if $m$ is odd} \\
\end{array}
\right.
\end{equation}
or
\begin{equation}  \label{sol3ccnew}
u_m =
\left\{
\begin{array}{l}
 C_1     \coth ( \psi  m  + C_2 ) + C_3  \qquad \mbox{if $m$ is even} \\
 \\
 C_1     \tanh ( \psi  m  + C_2 ) + C_3  \qquad \mbox{if $m$ is odd} \\
\end{array}
\right.
\end{equation}
or
\begin{equation}  \label{sol3dd}
u_m  = C_1  \mu_{1,2} ^m  + C_2
     = C_1   ( -1 ) ^m    e^{\pm  2 \psi m } + C_2
\end{equation}
on the mesh
\begin{equation}   \label{sol3ee}
x_m =
\left\{
\begin{array}{l}
 C_4     \tanh ( \psi  m  + C_5 ) + C_6  \qquad \mbox{if $m$ is even} \\
 \\
 C_4     \coth ( \psi  m  + C_5 ) + C_6  \qquad \mbox{if $m$ is odd} \\
\end{array}
\right.
\end{equation}
or
\begin{equation}   \label{sol3eenew}
x_m =
\left\{
\begin{array}{l}
 C_1     \coth ( \psi  m  + C_2 ) + C_3  \qquad \mbox{if $m$ is even} \\
 \\
 C_1     \tanh ( \psi  m  + C_2 ) + C_3  \qquad \mbox{if $m$ is odd} \\
\end{array}
\right.
\end{equation}
or
\begin{equation}   \label{sol3ff}
x_m  = C_4  \mu_{1,2} ^m  + C_5
     =  C_4   ( -1 ) ^m    e^{\pm  2 \psi m } + C_5  ,
\end{equation}
where $C_1 \neq 0$,  $C_2 $,   $C_3$, $C_4 \neq 0$,  $C_5  $ and $C_6$
are constants.
Here
$$
\mu_{1,2} =  { K - 2   \pm   \sqrt{K^2 - 4K} \over 2 }
$$
and
$$
\psi
=   {1\over 2} \ln ( - \mu_{1} )
=   {1\over 2} \ln \left( - { K - 2   +  \sqrt{K^2 - 4K} \over 2 } \right)  .
$$

\item

$ 0 < K <4 $

We obtain the solution
\begin{equation}  \label{sol3k}
u_m =  C_1  \tan ( \phi  m  + C_2 ) + C_3
\end{equation}
on the mesh
\begin{equation}  \label{sol3m}
x_m = C_4  \tan ( \phi m  + C_5 ) + C_6
\end{equation}
where $C_1 \neq 0$,  $C_2 $,   $C_3$, $C_4 \neq 0$,  $C_5  $ and $C_6$
are constants.
Here
$$
 \phi = \arccos \left( {\sqrt{K} \over 2}  \right) .
$$
\end{enumerate}

\begin{remark}
Let us note that all these solutions
can be presented in the unified form
\begin{equation}
u_m = { 1 \over \alpha x_m + \beta } + \gamma
\qquad
\mbox{or}
\qquad
u_m = \alpha x_m  + \beta ,
\end{equation}
where $\alpha \neq 0  $, $\beta $  and $ \gamma$ are constants.
They should be considered on the corresponding meshes,
which are different for different values of the parameter $K$.
Thus, the discretization~(\ref{discrete3e}) provides
the exact solution of the ODE~(\ref{odePW}) for $M= 0$.
For the case $K = 4$ this was observed in~\cite{[Pavel], [Pavel2]}.
\end{remark}

It should be noted that for $K < 0 $
we do not get monotonicity for mesh points $x_m$.
This is clearly seen in mesh
equations~(\ref{sol3ee}), (\ref{sol3eenew}) and~(\ref{sol3ff}).
Though we obtain the exact solution of the ODE in these points,
we can not speak about a mesh
on which we have a discretization of the ODE.

\eject


\bigskip

\noindent {\bf Case: $M \neq  0 $}

\medskip

As a discretization we consider the invariant scheme
\begin{equation}   \label{discrete3f}
\begin{array}{c}
{\displaystyle
F =  {  ( u_{m+3} -  u_{m+1} ) (  u_{m+2} -  u_{m} )   \over
             ( u_{m+3} -  u_{m+2} ) (  u_{m+1} -  u_{m} )  }
- {  ( x_{m+3} -  x_{m+1} ) (  x_{m+2} -  x_{m} )   \over
             ( x_{m+3} -  x_{m+2} ) (  x_{m+1} -  x_{m} )  }  } \\
\\
{\displaystyle
\times
\left(
1 - { M \over 6}    ( x_{m+3} -  x_{m} ) (  x_{m+2} -  x_{m+1} )
\right)       = 0 , } \\
\\
{\displaystyle
\Omega (   x_{m+3} -  x_{m+2} ,    x_{m+2} -  x_{m+1} ,  x_{m+1} -  x_{m})  = 0    .   } \\
\end{array}
\end{equation}
It admits the four symmetries~(\ref{part1}).
To find solutions we specify the mesh as a regular one
\begin{equation}   \label{discrete3ff}
 \Omega =   x_{m+1} -  x_{m} - h   = 0   ,
\end{equation}
where $  h >  0  $ is a constant.
The first equation will take the form
\begin{equation}    \label{modified}
F =  {  ( u_{m+3} -  u_{m+1} ) (  u_{m+2} -  u_{m} )   \over
             ( u_{m+3} -  u_{m+2} ) (  u_{m+1} -  u_{m} )  }
- \bar{K} = 0 ,
\end{equation}
where
$$
\bar{K} = 4  \left(   1 - { M \over 2} h^2 \right) .
$$
For the equation~(\ref{modified}) we can use results
obtained for the mapping~(\ref{discrete2e}) in Paragraph~\ref{four-point}.
Since $  h \neq 0  $ we have $\bar{K} \neq 4$.
For nontrivial cases $\bar{K} \neq 0$
there can be three possibilities.

\begin{enumerate}

\item

$ 0 < \bar{K} < 4$ ($M > 0$, $ 0 < h < \sqrt{2 / M} $)

We obtain the solution
\begin{equation}
u_m =  C_1  \tan ( \bar{\phi}  m  + C_2 ) + C_3 ,
\end{equation}
where $C_1 \neq 0$,  $C_2 $ and $C_3$ are constants,
on the mesh
\begin{equation}  \label{regularmesh}
x_m =  x_0  + h m  .
\end{equation}
Here
$$
\bar{\phi} = \arccos \left( { \sqrt{\bar{K}} \over 2 } \right)  .
$$

\item

$  \bar{K}> 4$ ($M < 0$)

We obtain the solution
\begin{equation}
u_m =  C_1   \tanh ( \bar{\psi} m  + C_2  ) + C_3
\end{equation}
or
\begin{equation}
u_m =  C_1   \coth ( \bar{\psi} m  + C_2  ) + C_3
\end{equation}
or
\begin{equation}
u_m  = C_1  \bar{\mu}_{1,2} ^m  + C_2
     = C_1  e^{ \pm 2 \bar{\psi} m }   + C_2 ,
\end{equation}
where $C_1 \neq 0$,  $C_2  $ and  $C_3$ are constants,
on the regular mesh~(\ref{regularmesh}).
Here
$$
\bar{\mu}_{1,2} =  { \bar{K} - 2   \pm   \sqrt{\bar{K}^2 - 4\bar{K}} \over 2 }
$$
and
$$
\bar{\psi}
=   { 1 \over 2 }   \ln \bar{\mu}_{1}
=   { 1 \over 2 }   \ln \left( { \bar{K} - 2   + \sqrt{ \bar{K}^2 - 4 \bar{K} } \over 2 } \right)  .
$$

\item

$  \bar{K} < 0 $  ($M > 0$, $h > \sqrt{2 / M} $)

We obtain the solution
\begin{equation}
u_m =  \left\{
\begin{array}{l}
 C_1     \tanh ( \bar{\psi}  m  + C_2 ) + C_3  \qquad \mbox{if $m$ is even} \\
 \\
 C_1     \coth ( \bar{\psi}  m  + C_2 ) + C_3  \qquad \mbox{if $m$ is odd} \\
\end{array}
\right.
\end{equation}
or
\begin{equation}
u_m = \left\{
\begin{array}{l}
 C_1     \coth ( \bar{\psi}  m  + C_2 ) + C_3  \qquad \mbox{if $m$ is even} \\
 \\
 C_1     \tanh ( \bar{\psi}  m  + C_2 ) + C_3  \qquad \mbox{if $m$ is odd} \\
\end{array}
\right.
\end{equation}
or
\begin{equation}
u_m  = C_1  \bar{\mu}_{1,2} ^m  + C_2
     = C_1  (-1) ^m  e^{ \pm 2 \bar{\psi} m }   + C_2 ,
\end{equation}
where $C_1 \neq 0$,  $C_2  $ and  $C_3$ are constants,
on the regular mesh~(\ref{regularmesh}).
Here
$$
\bar{\mu}_{1,2} =  { \bar{K} - 2   \pm   \sqrt{\bar{K}^2 - 4\bar{K}} \over 2 }  .
$$
and
$$
\bar{\psi}
=   { 1 \over 2 }   \ln ( - \bar{\mu}_{1}  )
=   { 1 \over 2 }   \ln \left( - { \bar{K} - 2   + \sqrt{ \bar{K}^2 - 4 \bar{K} } \over 2 } \right) .
$$

Note that because of the steplength  restriction $h > \sqrt{2 / M} $
we can not speak about consistent discretization of the ODE in this case.

\end{enumerate}

It should be noted that for sufficiently small steplengths
$h \ll 1$ we will always have  $\bar{K} > 0 $
and, thus, avoid jumping solutions of the last case.

\begin{remark}
We recall that in the case $M = 0$ the scheme~(\ref{discrete3e})
provided us with the exact solution of the ODE~(\ref{odePW}).
In the present case   $M \neq 0$
the scheme~(\ref{discrete3f}) with regular mesh  specification~(\ref{discrete3ff})
provides the exact solutions of the ODE~(\ref{odePW})
if we apply the scheme to the modified equation
\begin{equation}     \label{newode}
 F  _{mod} =  { 1 \over \dot{u} ^2 } \left( \dot{u}  \dddot{u} - { 3 \over 2 }  \ddot{u} ^2  \right) - M_{mod}  = 0 .
\end{equation}
The original equation parameter $ M > 0 $ should be changed to
the modified value
$$
 M_{mod}  = { 2 \over h^2 }  \sin ^2 \left( \sqrt{M \over 2} h \right),
 \qquad
0 < h < \sqrt{2 / M}
$$
and the parameter $ M < 0 $ should be changed to
the modified value
$$
 M_{mod}  = - { 2 \over h^2 }  \sinh ^2 \left( \sqrt{-{M \over 2}} h \right) .
$$
Application of the scheme~(\ref{discrete3f})
to the modified equation~(\ref{newode}) with the modified constant $ M_{mod} $
gives exact solution of the ODE~(\ref{odePW}) with constant $M$.
Note that in both cases $ M_{mod} \rightarrow M$  as $ h  \rightarrow 0 $.

We note that modification of the constant $M$ can be interpreted
as  scaling of the independent variable $x$.
\end{remark}

\section{Conclusion}

\label{conclusion}


This paper consists of two parts. The first is a brief review of two known
methods of obtaining first integrals of differential equations with
nontrivial Lie symmetries: the "adjoint equation method"
(Sections~\ref{adjointsection} and~\ref{sectionODE})
and the "direct method" (Section~\ref{integrating}).
Both of them are particularly useful either when
no Lagrangian exists, or when the symmetries of the equation are not
Lagrangian ones and the Noether theorem can not be applied. The methods
are valid both for ordinary and partial differential equations. We apply
them to obtain first integrals and general solutions of a third order
nonlinear ODE (the Schwarzian equation~(\ref{third})).

The second part is an adaptation of the adjoint equation method first to
mappings, then to difference systems. The mappings are equations involving
several discrete points. The second are difference equations on lattices
that arise e.g. when differential equations are solved numerically.
In  both cases (see Sections~\ref{discrete} and~\ref{discretization}, respectively)
we apply the discretized adjoint equation method to a specific four-point
equation, respectively four-point difference systems. These systems have
the Schwarzian ODE~(\ref{third}) as a continuous limit and and share its Lie
point symmetry group. We have also treated a simpler example, namely a
discrete linear harmonic oscillator.
The results for  both examples can be summed up as follows:

\begin{enumerate}

\item

The adjoint equation method makes it possible to obtain complete sets
of functionally independent first integrals of the differential equations and
difference systems. These in turn provide the general solutions of these
equations.

\item

The invariant discretization of continuous ODE~(\ref{third}) with $M = 0 $
considered here is exact.
The solutions of the difference system coincide with the solutions
of the original ODE.
The invariant discretizations of the other continuous ODEs considered here, namely
of the harmonic oscillator and  ODE~(\ref{third}) with $ M \neq 0$,
can be made exact if we allow a parameter modification.

\end{enumerate}

In the paper we restricted ourselves to ordinary difference equations.
However, the presented approach can be extended to differential--difference equations
as well as to partial difference equations.


\begin{thebibliography}{99}

\label{refereses}


\par
\bibitem{[AthertonHomsy]}
R.~W.~Atherton and G.~M.~Homsy
(1975)
On the existence and formulation of variational principles for nonlinear differential equations
{\it Stud. Appl. Math.} {\bf 54} (1)  31--60.



\par
\bibitem{[Bluman1]}
S.~Anco and G.~Bluman
(1997)
Direct construction of conservation laws from field equations,
{\it Phys. Rev. Lett.} {\bf 78}  2869--2873.




\par
\bibitem{[Blu]}
G.~Bluman and S.~Anco
(2002)
{\it Symmetry and Integration Methods for Differential Equations}
(New York: Springer)




\par
\bibitem{[Bluman2]}
G.~Bluman, A.~Cheviakov and S.~Anco
(2010)
{\it Applications of Symmetry Methods to Partial Differential Equations},
Vo. 168, Appl. Math. Sci.
(New York: Springer)


\par
\bibitem{[Pavel]}
A.~Bourlioux, C.~Cyr-Gagnon and P.~Winternitz
(2006)
Difference schemes with point symmetries and their numerical tests
{\it J. Phys A: Math. Gen.}  {\bf 39} (22)  6877--6896.



\par
\bibitem{[Pavel2]}
A.~Bourlioux, R.~Rebelo and P.~Winternitz
(2008)
Symmetry preserving discretization of $SL(2, \mathbb{R})$ invariant equations
{\it J. Nonlinear Math. Phys} {\bf 15}  362--372.




\par
\bibitem{[Dem]}
R.~Dennemeyer
(1968)
{\it Introduction to partial differential equations and boundary value problems}
(New York: McGraw-Hill)


\par
\bibitem{[Dor_3]}
V.~A.~Dorodnitsyn
(1991)
Transformation groups in mesh spaces
{\it J. Sov. Math.}  {\bf 55}  N1   1490,
Plenum Publishing Corporation.


\par
\bibitem{[Dor_4]}
V.~A.~Dorodnitsyn
(1993)
Finite--difference  models  entirely inheriting symmetry of original differential equations
{\it Modern Group Analysis:
Advanced Analytical and Computational Methods  in Mathematical Physics}  191,
Kluwer Academic Publishers.

\par
\bibitem{[Dor_5]}
V.~A.~Dorodnitsyn
(1993)
The finite-difference analogy of Noether's theorem
{\it Doklady RAN} {\bf 328} (6) 678--682 (in Russian).
Translation in {\it Phys. Dokl.} {\bf 38}  (2)  66--68  (1993).


\par
\bibitem{[66]}
V.~Dorodnitsyn
(2001)
Noether--type theorems for difference equations,
{\it Applied Numerical Mathematics}  {\bf 39}     307--321.



\par
\bibitem{[D-book]}
V.~Dorodnitsyn
(2011)
{\it Applications of Lie Groups to Difference Equations}
Chapman \& Hall/CRC differential and integral equations series.





\par
\bibitem{[DKap]}
V.~Dorodnitsyn and E.~Kaptsov
(2013)
Discretizing of second order ODEs possesing symmetries
{\it Journ. of computational mathematics and mathematical physics}
{\bf 53} (8)  1329--1355.



\par
\bibitem{[Dheat]}
V.~Dorodnitsyn and R.~Kozlov
(2003)
A heat transfer with a source: the complete set of invariant difference schemes
{\it  J. Nonlinear Math. Phys.}  {\bf 10}  (1)  16--50.



\par
\bibitem{[DK-1]}
V.~Dorodnitsyn and R.~Kozlov
(2009)
First integrals of difference  Hamiltonian equations
{\it J. Phys. A: Math. Theor.} {\bf  42}  454007.


\par
\bibitem{[DK-2]}
V.~Dorodnitsyn and R.~Kozlov
(2010)
Invariance and first integrals of continuous and discrete  Hamiltonian equations
{\it  J. Engrg. Math.} {\bf  66}  253--270.



\par
\bibitem{[DK-3]}
V.~Dorodnitsyn and R.~Kozlov
(2011)
Lagrangian and Hamiltonian formalism for discrete equations: symmetries and first integrals,
SMS Lecture Notes, SMS Lecture Notes, Cambridge University Press, 7--49.



\par
\bibitem{[D_K_W_1]}
V.~Dorodnitsyn, R.~Kozlov and P.~Winternitz
(2000)
Lie group classification of second-order ordinary difference equations,
{\it J. Math. Phys.}  {\bf 41} (1)  480--504.



\par
\bibitem{[45]}
V.~Dorodnitsyn, R.~Kozlov and P.~Winternitz
(2004)
Continuous symmetries of Lagrangians and exact solutions of discrete equations
{\it J. Math. Phys.}  {\bf 45} (1)  336--359.






\par
\bibitem{[Hydon-Mansfield]}
P.~E.~Hydon
(2000)
Symmetries and first integrals of ordinary difference equations
{\it  R. Soc. Lond. Proc. Ser. A Math. Phys. Eng. Sci. }
{\bf 456} (2004)  2835--2855.



\par
\bibitem{[Ibr1]}
N.~H.~Ibragimov
(1985)
{\it Transformation Groups Applied to Mathematical Physics}
(Dordrecht:  Reidel)





\par
\bibitem{[Ibr2]}
N.~Ibragimov
(2007)
A new conservation theorem
{\it J. Math. Anal. Appl.}
{\bf  333}, 311--328.


\par
\bibitem{[ibr11a]}
N.~H.~Ibragimov
(2011)
Nonlinear self-adjointness  and conservation laws
{\it J. Phys A: Math. Gen.} {\bf 44}  432002.


\par
\bibitem{[Ibr]}
N.~Ibragimov
(2010--2011)
Nonlinear self-adjointness in constructing conservation laws,
Archives of ALGA {\bf 7/8}, ALGA Publications, Karlskrona, Sweden.




\par
\bibitem{[KaraMahomed]}
A.~H.~Kara and F.~M.~Mahomed
(2000)
Relationship between symmetries and conservation laws
{\it Internat. J. Theoret. Phys.}  {\bf  39}  (1)   23--40.




\par
\bibitem{[LW-4]}
D.~Levi and P.~Winternitz
(1991)
Continuous symmetries of discrete equations
{\it  Phys. Lett. A}  {\bf 152} (7) 335--338.


\par
\bibitem{[LW-5]}
D.~Levi and P.~Winternitz
(1996)
Symmetries of discrete dynamical systems
{\it J. Math. Phys.} {\bf 37} 5551--5576.


\par
\bibitem{[LW-2]}
D.~Levi and P.~Winternitz
(2006)
Continuous symmetries of difference equations
{\it  J.Phys. A} {\bf 39} (2) R1--R63.




\par
\bibitem{[LW-6]}
D.~Levi, P.~Winternitz and R.~I.~Yamilov.
(2010)
Lie point symmetries of differential--difference equations
{\it  J. Phys. A Math. Theor.}
{\bf 43} (29) 292002.


\par
\bibitem{[LW-11]}
D.~Levi, Z.~Thomova and P.~Winternitz,
(2011)
Are there contact transformations for discrete equations?
{\it  J. Phys. A: Math. Theor.}  {\bf 44}  265201.



\par
\bibitem{[LW-12]}
D.~Levi, C.~Scimiterna, Z.~Thomova and P.~Winternitz
(2012)
Contact transformations for difference schemes
{\it  J. Phys. A: Math. Theor. } {\bf  45}  022001.



\par
\bibitem{[Lut]}
M.~Lutzky
(1978)
Symmetry groups and conserved quantities for the harmonic oscillator
{\it J. Phys. A}  {\bf 11} (2)  249--258.



\par
\bibitem{[Noe1]}
E.~Noether
(1918)
Invariante Variationsprobleme,
Nachr. Konig. Gesell. Wissen., Gottingen,
{\it Math.-Phys. Kl.}  {\bf 2}  235--257.




\par
\bibitem{[Olver1]}
P.~J.~Olver
(1993)
{\it Applications of Lie groups to differential equations}
Second edition
(New York: Springer--Verlag)


\par
\bibitem{[Olver-1]}
P.~J.~Olver
(2001)
Geometric foundations of numerical algorithms and symmetry
{\it  Appl. Alg. Engin. Comp. Commun.}
{\bf 11}  417--436.

\par
\bibitem{[Ovsienko]}
V.~Ovsienko and S.~Tabachnikov
(2009)
What is the Schwarzian derivative?
{\it Notices of the AMS}  {\bf 56} (1)  34--36.


\par
\bibitem{[Ovs1]}
L.~V.~Ovsyannikov
(1982)
{\it Group analysis of differential equations}
(New York: Academic)



\par
\bibitem{[Quispel-1]}
G.~R.~W.~Quispel, H.~W.~Capel and R.~Sahadevan
(1992)
Continuous symmetries of differential--difference equations
{\it Phys. Lett. } {\bf 170A} 379--383.






\par
\bibitem{[Rosenhaus1]}
V.~Rosenhaus and  G.~H.~Katzin
(1994)
On symmetries, conservation laws, and variational
problems for partial differential equations
{\it J. Math. Phys. } {\bf 35 } (4) 1998--2012.









\par
\bibitem{[LW-3]}
P.~Winternitz
(2011)
Symmetry preserving discretization of differential equations
and Lie point symmetries of differential-difference equations.
In D.~Levi, P.~J.~Olver, Z.~Thomova and P.~Winternitz, editors,
Symmetries and Integrability of Difference Equations,
292--341. Cambridge University Press.




\par
\bibitem{[DAN]}
P.~Winternitz, V.~Dorodnitsyn, E.~Kaptsov and R.~Kozlov,
First integrals of difference equations
which do not possess a variational formulation (in Russian),
{\it Doklady Mathematics} (To appear);
arXiv:1307.7585 (10 pages).





\end{thebibliography}
\end{document}